\documentclass[journal,10pt]{IEEEtran}

\usepackage{amssymb}
\usepackage{amsmath}
\usepackage{cite}
\usepackage{url}
\usepackage{xcolor}
\usepackage{cite,graphicx,amsmath,amssymb}
\usepackage{subfigure}
\usepackage{citesort}
\usepackage{fancyhdr}
\usepackage{mdwmath}
\usepackage{mdwtab}
\usepackage{caption}
\usepackage{amsthm}

\newtheorem{remark}{Remark}
\newtheorem{theorem}{Theorem}

\newtheorem{lemma}{Lemma}

\newtheorem{corollary}{Corollary}

\newtheorem{proposition}{Proposition}
\newtheorem{assumption}{Assumption}

\makeatletter
\def\ScaleIfNeeded{%
\ifdim\Gin@nat@width>\linewidth \linewidth \else \Gin@nat@width
\fi } \makeatother

\DeclareMathOperator\erfc{erfc}

\begin{document}

\title{Cache-enabled HetNets With Millimeter Wave Small Cells}

\author{Wenqiang~Yi,~\IEEEmembership{Student Member,~IEEE,}
        Yuanwei~Liu,~\IEEEmembership{Member,~IEEE,}
        and Arumugam~Nallanathan,~\IEEEmembership{Fellow,~IEEE,}
\thanks{W. Yi, Y. Liu, and A. Nallanathan are with Queen Mary University of London, London, UK (email: \{w.yi, yuanwei.liu, a.nallanathan\}@qmul.ac.uk).}
\thanks{Part of this work was presented in IEEE International Conference on Communications (ICC), May, USA, 2018~\cite{123456789}.}
}

\maketitle

\begin{abstract}
  In this paper, we consider a novel cache-enabled heterogeneous network (HetNet), where macro base stations (BSs) with traditional sub-6 GHz are overlaid by dense millimeter wave (mmWave) pico BSs. These two-tier BSs, which are modeled as two independent homogeneous Poisson Point Processes, cache multimedia contents following the popularity rank. High-capacity backhauls are utilized between macro BSs and the core server. In contrast to the simplified flat-top antenna pattern analyzed in previous articles, we employ an actual antenna model with the uniform linear array at all mmWave BSs. To evaluate the performance of our system, we introduce two distinctive user association strategies: 1)~maximum received power (Max-RP) scheme; and 2)~maximum rate (Max-Rate) scheme. With the aid of these two schemes, we deduce new theoretical equations for success probabilities and area spectral efficiencies~(ASEs). Considering a special case with practical path loss laws, several closed-form expressions for coverage probabilities are derived to gain several insights. Monte Carlo simulations are presented to verify the analytical conclusions. We show that: 1) the proposed HetNet is an interference-limited system and it outperforms the traditional HetNets in terms of the success probability; 2) there exists an optimal pre-decided rate threshold that contributes to the maximum ASE; and 3) Max-Rate achieves higher success probability and ASE than Max-RP but it needs the extra information of the interference effect.
\end{abstract}

\begin{IEEEkeywords}
Caching, heterogeneous networks, millimeter wave, stochastic geometry, user association
\end{IEEEkeywords}

\section{Introduction}

  With the rapid development of the traditional cellular networks and novel Internet-enabled applications, such as multimedia sensors~\cite{7835615} and electric vehicles~\cite{7909161,7926936}, the total throughput of mobile networks in 2020 is expected to become 1000-fold larger than that in 2010~\cite{cisco2014global}. To support the explosive data traffic of future fifth-generation (5G) cellular networks, numerous researches~\cite{bhushan2014network,7954630,7982794,Zhiguo2015Mag} have paid attention to an innovative framework that densifies the traditional networks with massive small base stations (BSs). However, the improvement of these heterogeneous networks (HetNets) is mainly restricted to the capacity of the backhauls. Although the high-speed optical fiber provides a theoretical solution, in practice, connecting the core server to all BSs with fibers is arduous and costly~\cite{liu2017caching}. Moreover, microwave backhauls may pessimistically weaken the throughput gain fetched by the network densification~\cite{4623708}. A recent study~\cite{cha2007tube} has shown that only 5\%-10\% of multimedia contents are required by the majority of user equipments (UEs). Additionally, the storage capacity of cache-enabled devices expands promptly at a fairly low cost. Stimulated by such facts, equipping caches at all BSs for storing the most popular contents becomes a promising method to offload the data traffic rather than continuing increasing the networks' density~\cite{6495773,6736753}.

  Lately, the aforementioned cache-enabled HetNets have been studied in various papers. Authors in~\cite{7445129} analyzed the energy efficiency and throughput of cellular networks with caches, but they only considered the small cell networks (SCNs) and BSs were modeled following a regular hexagonal grid. Since stochastic geometry is a useful tool to acquire the networks' randomness~\cite{7445146}, modeling a tier of BSs in SCNs or HetNets with a homogeneous Poisson Point Process~(HPPP) is more accurate than the traditional hexagonal scenario~\cite{jo2012heterogeneous,elsawy2013stochastic,heath2013modeling}. Under this condition, the throughput of multi-tier cache-enabled HetNets was discussed in~\cite{yang2016analysis}, where BSs were modeled as mutually independent PPPs. However, the high-capacity backhauls were employed at all nodes including the macro BSs and relays, which is uneconomical in reality. Then the limitation was relaxed by assuming that only macro BSs connected the core networks through backhauls, while BSs in small cells cached the contents via wireless broadcasting~\cite{liu2017caching}. Unfortunately, the further analysis on the impact of backhaul capacity was omitted, which is the key parameter when comparing with the conventional HetNets.

  In addition to the network densification, another key capacity-increasing technology for boosting the throughput of future cellular networks is exploiting new spectrum bands, such as millimeter wave (mmWave)~\cite{andrews2014will,boccardi2014five,8114722}. Comparing with the traditional sub-6 GHz networks in 4G, two distinctive characteristics of mmWave are small wavelength and the sensitivity to blockages~\cite{6932503}. Thanks to the short wavelength, steerable antennas with huge scales can be employed at devices to enhance the directional array gain~\cite{rappaport2013broadband}. On the other side, the sensitivity gives rise to severe penetration loss for mmWave signals when passing through building exteriors~\cite{alejos2008measurement}. Therefore, the path loss law of non-line-of-sight (NLOS) links is substantially different from that of line-of-sight (LOS) links in mmWave communications~\cite{rappaport2013broadband,6515173}, and it is unrealistic to expect an outdoor-to-indoor coverage from macro mmWave BSs. To compensate the blockage-dependent loss, an ingenious hybrid network is created, where mmWave transmitters contribute to the ultra-fast data rate in short-range small cells, and sub-6~GHz BSs provide the universal coverage~\cite{7493676}.

  There exist numerous studies concentrating on the performance of mmWave communications. As discussed in cache-enabled HetNets, stochastic geometry has also been widely utilized in mmWave networks, where the locations of transceivers were modeled following PPPs~\cite{6932503,andrews2011tractable}. With the aid of such structure, the primary article~\cite{6932503}, which employed a simplified flat-top antenna pattern, introduced a stochastic blockage model to represent the actual mmWave communication environment. In fact, this simplified model has limited ability to exactly depict several parameters of a practical antenna, such as beamwidth, front-back ratio and nulls~\cite{170204493}. Therefore, the authors in~\cite{maamari2016coverage} proposed an actual antenna pattern for traditional mmWave networks. Considering the hybrid HetNets, a tractable framework with sub-6~GHz macro cells and mmWave small cells was analyzed under two user association strategies in~\cite{7493676}. However, the Rayleigh fading assumption is not accurate for mmWave communications because of the poor scattering feature~\cite{170204493}. Recent works~\cite{6932503,8016632} presented a realistic channel model with Nakagami fading to improve the theoretical accuracy.

 \subsection{Motivation and Contribution}

  Although HetNets with caches have been analyzed under a variety of scenarios with traditional sub-6 GHz networks, there is still lack of articles on a hybrid system with mmWave small cells. Since mmWave has a large range of available bandwidth~\cite{rappaport2014millimeter,pi2011introduction} and it is able to provide fast data rate in short-distance networks~\cite{park2007short}, adopting mmWave into a dense pico tier of HetNets is a promising way to increase the throughput of 5G cellular networks. Additionally, utilizing low-cost caches at all macro and pico BSs is capable of offloading the backhaul traffic efficiently and hence providing further improvement regarding the quality of service. The other benefit of such hybrid HetNets is no mutual interferences because each tier uses totally distinctive carrier frequency. These advantages motivate us to create this paper.

  In contrast to~\cite{liu2017caching}, we introduce fiber-connections between macro BSs and the multimedia server to evaluate the impact of backhaul capacity in cache-enabled HetNets. Then, due to the employment of mmWave, the propagation environment and antenna beamforming pattern in the small cells are replaced by Nakagami fading and actual antenna arrays, respectively. Load balancing problems in mmWave-enabled HetNets have been studied in~\cite{7841752,6774981}. However, the optimal solutions are based on a simplified framework which ignores the randomness of network nodes. In order to enhance the generality, we use the stochastic geometry to model the locations of transceivers. Regarding the user association scheme in hybrid HetNets (mmWave plus sub-6~GHz), authors in~\cite{6497002} assumed that the typical user is served by the BS which offers the minimum path loss. In most instances, even mmWave transmissions have more severe path loss than sub-6~GHz scenarios, they are still able to provide faster data rate due to the huge transmit bandwidth. As a result, the maximum data rate is also an important criterion for user association, in addition to the minimum path loss. We consider both criteria in this paper. The main contributions are summarized as below:

\begin{itemize}
  \item The success probability and area spectral efficiency (ASE) of our hybrid cache-enabled HetNets are discussed under two user association strategies: 1) \emph{Maximum Received Power~(Max-RP)}, where the requesting user chooses the macro BS or pico BS offering the maximum average biased received power\footnote{The motivation for considering the average received power is that the network designer is interested in the average metric at the requested user for the universal coordination~\cite{6171996}.} from all BSs containing the requested file; and 2) \emph{Maximum Rate~(Max-Rate)}, where the typical UE selects the BS, which provides the highest biased transmitting rate, from all BSs caching the desired content.
  \item We analyze the cache-related coverage performance of traditional sub-6 GHz macro cells and mmWave small cells with the actual antenna pattern. Furthermore, closed-form coverage probability equations for the mmWave tier and an interference-limited case with sub-6 GHz are derived. Our analytical expressions can be directly applied into other mmWave or sub-6~GHz scenarios with negligible changes.
  \item Different association probabilities for two considered schemes are introduced to calculate final algorithms of success probabilities. We theoretically demonstrate that the success probability of Max-RP has a positive correlation with the serving tier's biased transmit power. However, the success probability of Max-Rate scheme is independent of the two tier's transmit power and the density of macro BSs. Finally, expressions of ASEs are deduced for analyzing.
  \item  We conclude that: 1) our cache-enabled hybrid HetNets outperform the traditional HetNets where macro BSs have no caching capacity, and Max-Rate achieves a better performance than Max-RP in terms of the success probability and ASE; 2) the proposed network is an interference-limited system due to the nature of sub-6 GHz networks and the high density of mmWave small cells; 3) there is an optimum value of rate requirement for obtaining the maximum ASE; and 4) 73 GHz is the best mmWave carrier frequency for two user association strategies because of possessing the largest antenna scale.
\end{itemize}

\subsection{Organization}

  We organize the rest of our treatise as follows: In Section II, we present the system model where two-tier BSs and users in the proposed cache-enabled hybrid HetNets are modeled as three independent HPPPs. In Section III, the expressions of signal-to-interference-plus-noise-ratio~(SINR) coverage probabilities for two distinctive tiers are derived with the aid of the random content placement scheme. In Section IV, we discuss two different user association strategies, based on which the algorithms of success probabilities and ASEs are deduced. In Section V, the simulation and numerical results are presented for corroborating the analytical conclusions and providing further analysis, respectively. In Section VI, we draw our conclusions.

\section{System Model}\label{System Model}

\subsection{Network Architecture}

\begin{figure*} [ht!]
\centering
\includegraphics[height=2.2in]{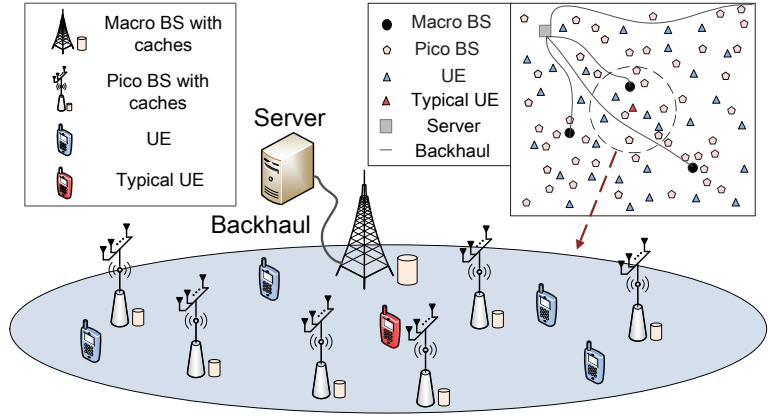}
\caption{Layouts of the proposed cache-enabled hybrid HetNet with traditional macro cells and mmWave small cells.}
\label{model}
\end{figure*}
In this paper, we present a cache-enabled hybrid HetNet with two-tier BSs as shown in Fig.~\ref{model}. Macro BSs, pico BSs, and UEs are distributed following three independent HPPPs with density $\lambda_1$, $\lambda_2$, and $\lambda_u$, denoted by $\Phi_1$, $\Phi_2$, and $\Phi_u$, respectively. A randomly selected \emph{typical UE} is fixed at the origin such that the probability density function~(PDF) of the distance from the typical UE to its nearest BS in the $i$-th tier is given by $p_i(r)=2\pi\lambda_i r \exp (-\pi \lambda_i r^2)$, where $i \in \{1,2\}$. Apparently, the number of pico BSs in real HetNets is much more than that of macro BSs and thus we consider $\lambda_2 \gg \lambda_1$. In order to compare the performance of the proposed network with traditional HetNets, we provide a server to supply the less-popular contents. Note that deploying wired connections between the core server and all pico BSs is wasted and arduous. We assume the server only connects to each macro BS through a high-capacity wired backhaul.

In order to avoid inter-tier interference, hybrid carrier frequencies are employed in our system. When communicating with UEs, the macro BSs adopt sub-6 GHz, while the pico BSs utilize mmWave. Note that various multiple-access techniques enable the macro BSs to serve multiple users in one time slot. We assume the quantity of UEs is large enough, namely $\lambda_u>\lambda_2 \gg\lambda_1$, to ensure all BSs are active when the typical UE is served.

\subsection{Blockage Model}

\begin{figure*}[t!]
\centering
\subfigure[Illustration of the LOS ball model utilized in the pico tier network.]{\label{LOS} \includegraphics[width= 2.6 in, height=2.1 in]{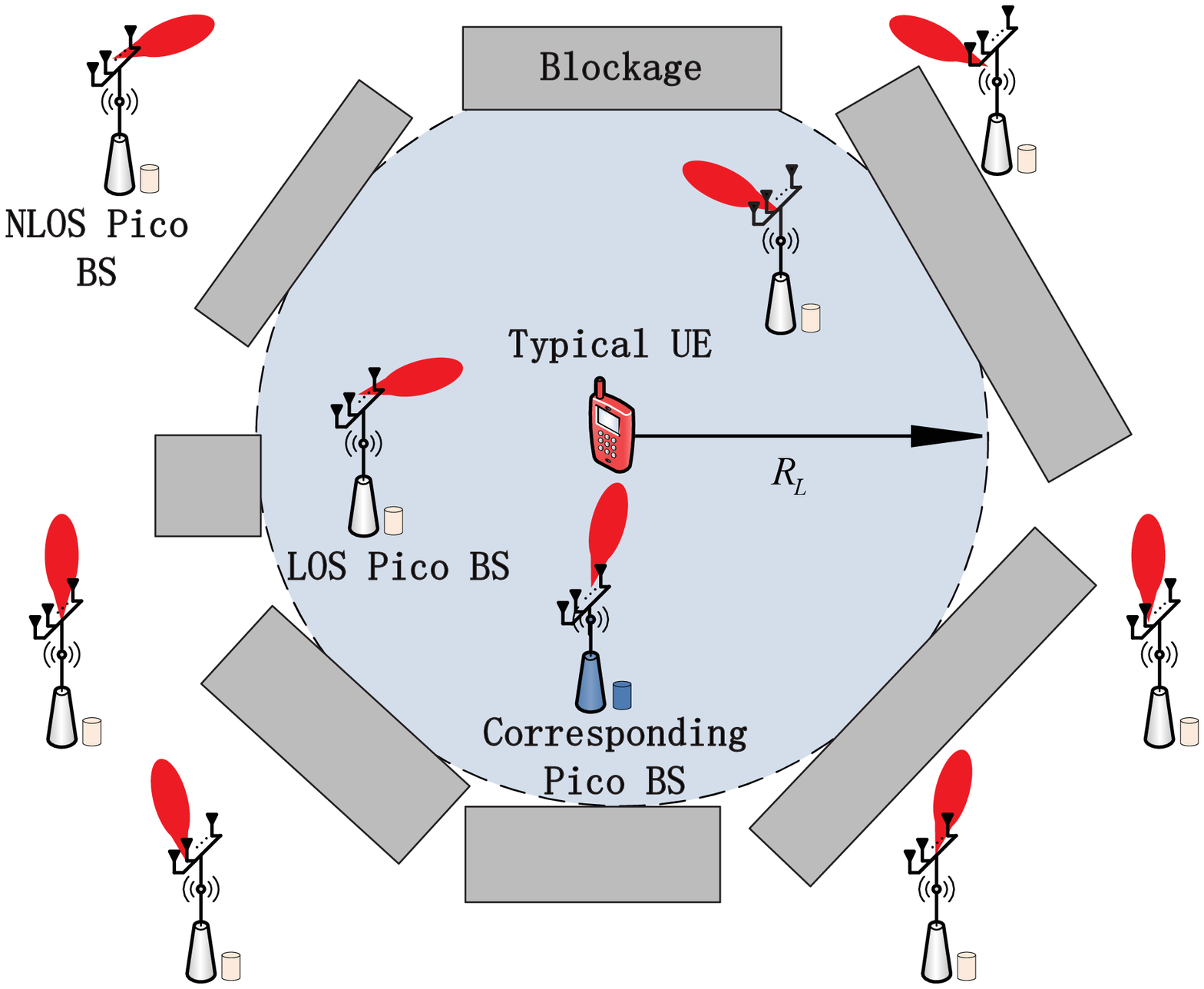}}
\subfigure[Illustration of two typical spatial AoDs for interfering mmWave BSs.]{\label{angle} \includegraphics[width= 1.9in, height=2.2 in]{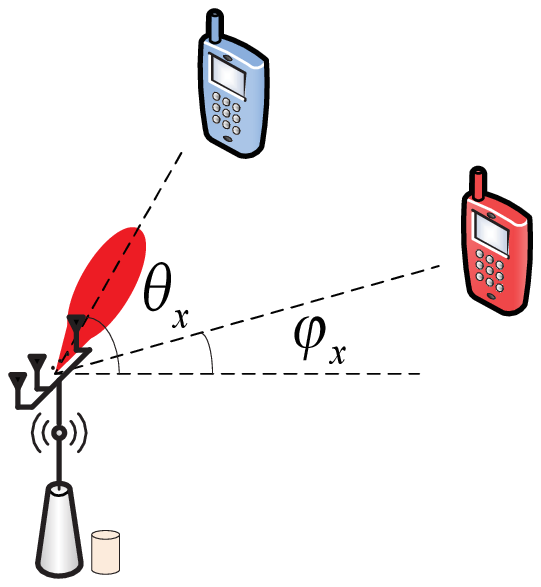}}
\caption{LOS Ball Model and Directional Beamforming in The Second Tier.}
\end{figure*}
In the first tier, when the communication distance is $r$, the path loss law $L_1(r)$ for sub-6~GHz signals is same as that in traditional cellular networks, which is given by
\begin{align}
L_1(r) = {{C_1}{r^{ - {\alpha _1}}},}
\end{align}
where $\alpha_1$ is the path loss exponent and $C_1$ is the intercept for the macro tier.

In the second tier, the effect of blockages is important due to the employment of mmWave. Therefore, we adopt a LOS ball model\footnote{In most urban scenarios, the considered LOS ball model has negligible deviation with the more accurate multi-slope LOS probability scheme, especially when the altitude of pico BSs is less than the average height of obstacles~\cite{LOSmodel}.}~\cite{6932503,7061455}, as shown in Fig.~\ref{LOS}, to depict the blockage environment. The radius $R_L$ for the LOS ball represents the departure from nearby obstacles. The probability of LOS links is one inside the ball and zero outside that area. A recent study~\cite{andrews2017modeling} has advocated that when the density of BSs is large, this blockage pattern has a negligible difference with the commonly used random shape theory model~\cite{7105406}. Note that we consider a dense pico tier. The simplified LOS ball model is capable of providing enough analytical accuracy. Regarding NLOS links, various articles~\cite{6932503,7448962} have demonstrated that the impact of NLOS signals can be ignored in mmWave networks due to their severe path loss. As a result, only LOS signals are analyzed in this paper\footnote{Regarding the interference via NLOS transmissions, it can still be ignored since the high density of pico BSs enhances not only the interference from NLOS links but also the counterpart from LOS links~\cite{6932503,8016632,7448962,7862785}.}. Accordingly, the path loss law in the second tier $L_2(r)$ can be expressed as follows
\begin{align}
{L_2}(r) = \mathbf{U}({R_L} -r){C_2}{r^{ - {\alpha _2}}},
\end{align}
where $\alpha_2$ and $C_2$ are the path loss exponent and the intercept of LOS links, respectively. $\mathbf{U}(x)$ is the unit step function, which is defined as
\begin{align}
\mathbf{U}(x) = \left\{ {\begin{array}{*{20}{c}}
   {1,} & {x \ge 0}  \\
   {0,} & {x < 0}  \\
\end{array}} \right..
\end{align}

\subsection{Cache-enabled Content Access Protocol}

In this paper, we assume that a static multimedia content catalog containing $N_c$ files is stored at the server and all files have the same size with $E$ bits. Each macro BS and pico BS have restricted storage capacities with $(M_1\times E)$ and $(M_2\times E)$ bits, respectively, which means the maximum storage capacity of the proposed HetNet obeys $M_c=\max(M_1,M_2)\leq N_c$. High-speed fiber backhauls are employed for connecting the core server to macro BSs like traditional HetNets. The backhaul capacity is $C_{bh}$. When the data traffic load becomes low, the contents are broadcast to all BSs following the random content placement scheme as discussed in~\cite{7502130} until the storage is fully occupied. On this basis, we introduce the requesting probability, content placement, association strategy, and access protocol in the following part.

\emph{Requesting Probability:} We apply the Zipf distribution to represent the probability of content being requested~\cite{6799322,cha2007tube,6805204}. If files are indexed according to the popularity, namely the first and the $N_c$-th files are the most and the least popular contents, respectively, the requesting probability of the $f$-th file $P_{f}$ is given by
\begin{align}\label{Zipf}
{P_{f}} = \frac{{{f^{ - \delta }}}}{{\sum\nolimits_{n = 1}^{{N_c}} {{n^{ - \delta }}} }},\ \ \ (1\leq f \leq N_c),
\end{align}
where $f$ is an integer and $\delta\geq0$ is the skew parameter of the popularity distribution.

\begin{figure} [t!]
  \centering
  \includegraphics[width= 3in]{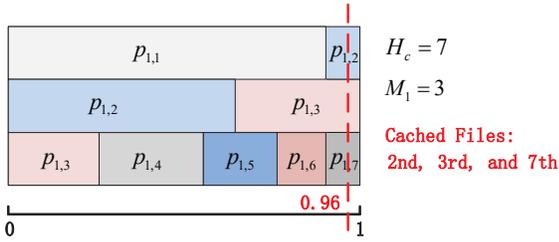}
   \caption{An example of content placement in the macro tier, with $M_1=3$ and $H_c=7$. The cache memory of one macro BS is uniformly divided into $M_1=3$ unit blocks (each row is one block). All blocks are sequentially filled with the probabilities from $p_{1,1}$ to $p_{1,H_c}$. After that, a random number between zero to one is selected to decide the cached files. In this example, the decision value is 0.96 such that the first, the third, and the 7-th files are cached in the macro tier.}
\label{content_placement}
\end{figure}

\emph{Content Placement:} We assume that the first $H_c$ files are cached in the considered HetNet. The probability that the $f$-th ranked file is cached at the $i$-th tier is denoted by $p_{i,f}$. Based on the optimal solution presented in~\cite{7248843}, such probability obeys:
\begin{align}
\sum\limits_{f = 1}^{H_c} {{p_{i,f}} = {M_i}} ,\ (M_c \le H_c \le N_c, 0 \le {p_{i,f}} \le 1).
\end{align}
An example in the macro tier is illustrated in Fig.~\ref{content_placement}. Note that the pico tier has the same content placement strategy but different storage capacities.

\emph{Association Strategy:} We consider association strategies depending on both cached files and channel conditions, which is essentially different with traditional HetNets without caches~\cite{liu2017caching}. In the $i$-th tier, the locations of BSs containing the $f$-th file $(1 \le f \le H_c)$ form a set $\Phi_{i,f}$ $(\Phi_{i,f} \subset \Phi_{i})$. When the typical UE requests this file, two association strategies are used in the considered HetNet: 1)~\emph{Max-RP}, where the typical UE communicates with the BS at $x_0$ $(x_0 \in \{\Phi_{i,f}\}_{i=1,2})$ that provides the maximum biased average received power; and 2)~\emph{Max-Rate}, where the typical UE connects to the BS at $x_0$ $(x_0 \in \{\Phi_{i,f}\}_{i=1,2})$ that provides the maximum biased received data rate.

\emph{Access Protocol:} If the desired $f$-th file is cached at the two-tier HetNet, which obeys $(1 \le f \le H_c)$, the typical UE communicates with the macro or pico BSs following the aforementioned association strategies. However, if the demanded content is absent from the BSs due to limited storage capacities, namely $(H_c < f \le N_c)$, the typical UE turns to the core server via the nearest macro BS.

\subsection{Directional Beamforming}

In the $i$-th tier, we employ antenna arrays composed of $N_i$ elements at all cache-enabled BSs and the transmit power is assumed to be a constant $P_i$. Due to the small wavelength of mmWave signals, the uniform linear array (ULA) pattern can be deployed at all pico BSs~\cite{170204493}. Directional antenna arrays deployed at the pico BSs supply substantial beamforming gains to compensate the path loss. However, we only consider an omnidirectional antenna model at macro BSs and UEs for tractability of the analysis~\cite{7493676}. When the typical UE requests the $f$-th file from the $i$-th tier, the received signal can be expressed as follows
\begin{align}\label{signaleq}
y_{i,f} =& \sqrt {P_iL_i(||x_0||)} {\textbf{h}^{{x_0}}_i}{\textbf{w}^{{x_0}}_i} {s_{{x_{0}}}} \nonumber \\
&+  \sum\limits_{{x} \in \Phi_{i} \backslash {x_{0}}} {\sqrt {P_iL_i(||x||)} {\textbf{h}^{{x}}_i}{\textbf{w}^{{x}}_i} {s_{{x}}}}  + \sigma_i,
\end{align}
where the typical UE is receiving the message from \emph{the corresponding BS} at $x_0$ $(x_0 \in \Phi_{i,f})$. The locations of all interfering BSs forms a set $\Phi_{i} \backslash {x_{0}}$ and each element in such set is donated by the variable $x$. The channel vector from the BS to the typical UE and the beamforming vector of the BS in the $i$-th tier are denoted by ${\textbf{h}^x_i}$ and ${\textbf{w}^x_i}$, respectively. $\sigma_i$ represents the thermal noise.

 Combining with the aforementioned assumptions, the product of the fading gain and beamforming gain of the BS located at $x$ in the $i$-th tier is shown as below~\cite{170204493}
\begin{align}
H^x_i\buildrel \Delta \over =|{\textbf{h}^x_i}{\textbf{w}^x_i}{|^2} = {N_i}|{h_i}{|^2}G_i( {{\varphi _x} - {\theta _x}} ),
\end{align}
where ${h_i}$ is the small fading term. ${\varphi _x} $ is the spatial angle of departure (AoD) from the interfering BS to the typical UE, and ${\theta _x}$ is the spatial AoD between the BS at location $x$ and its corresponding receiver, see Fig.~\ref{angle}. $G_i(.)$ is the array gain function. More specifically, the actual array pattern is employed at pico BSs so that $G_2( \omega  ) \buildrel \Delta \over = \frac{{{{\sin }^2}( {\pi {N_i}\omega } )}}{{N_i^2{{\sin }^2}( {\pi \omega } )}}$~\cite{170204493}, where $\omega$ is a uniformly distributed random variable over $[{ - \frac{d}{\lambda }},{ \frac{d}{\lambda }}]$. $d$ and $\lambda$ are the antenna spacing and wavelength, respectively~\cite{7279196}. On the other hand, the array gain function for macro BSs is $G_1( \omega  ) \buildrel \Delta \over = 1$ due to the omnidirectional antenna pattern.

Since sophisticated beam training protocols~\cite{rappaport2014millimeter} can be used at BSs to acquire the location information of the typical UE, we assume the corresponding BS provides the maximum directivity gain $G_0=1$ by aligning the antenna beam towards the typical UE.

\subsection{Propagation Model}

\subsubsection{Channel Model}
In the proposed HetNet, since the corresponding BS is interfered by other active BSs located in the same tier, the received SINR at the typical UE for requesting the $f$-th file from the $i$-th tier can be expressed as follows
\begin{align}
\Upsilon_{i,f}= \frac{{P_iL_i(||x_0||)G_0{N_i}|{h_i}{|^2}}}{{\sigma_i^2 + I_{i,f}+I_{i,f^\prime}}},
\end{align}
where $I_{i,f}=\sum\limits_{{x} \in {\Phi _{i,f}}\backslash {x_{0}}} {P_iL_i(||x||)H^{{x}}_i}$ and $I_{i,f^\prime}=\sum\limits_{{x} \in {\Phi _{i,f^\prime}}} {P_iL_i(||x||)H^{{x}}_i}$. $\Phi_{i,f^\prime}$ is the set of locations of BSs that do not store the $f$-th file and it obeys $\Phi_i=\Phi_{i,f}\bigcup\Phi_{i,f^\prime}$, $\Phi_{i,f}\bigcap\Phi_{i,f^\prime}={\O}$. $h_2$ follows independent Nakagami fading due to utilizing mmWave and the parameter of Nakagami fading $N^p_2$ is considered to be a positive integer for simplifying the analysis~\cite{6932503}. Therefore, $|h_2|^2$ is a normalized Gamma random variable. On the other side, we assume a Rayleigh fading model for the macro tier so that the fading parameter $N^p_1\buildrel \Delta \over = 1$.
\subsubsection{Association Criteria}
When the typical UE requests the $f$-th file. For Max-RP, the biased average received power is defined as follows~\cite{7982794}
\begin{align}
{{\bar P}_{i,f}} = b_i^P{P_i}{L_i}(||x_0||){N_i}{G_0},
\end{align}
where $b_i^P$ is a bias factor that aims to balance the load between two tiers under the Max-RP scheme~\cite{liu2017caching}. Then, we consider the biased received data rate for Max-Rate, which is given by
 \begin{align}
 R_{i,f} = b_i^B{B_i}{\log _2}\left( {1 + \Upsilon_{i,f}} \right),
 \end{align}
 where $B_i$ is the bandwidth per resource block. $b_i^B$ is another bias factor to control the data traffic under Max-Rate.

When the typical UE requests a file from the core server, the throughput can be limited by both the macro-tier conditions and the backhaul capacity~\cite{7445129}. Therefore, the instantaneous downlink data rate from the core server can be expressed as $R_1 = \min \left( {B_1{{\log }_2}\left( {1 + \Upsilon_{1,f}} \right),{C_{bh}}} \right).$

\section{Cache-Related SINR Coverage Probability}\label{Coverage_Probability}
Cache-related SINR coverage probability is the proportion of the received SINR that surpasses the requested SINR threshold $\tau \in \mathbb{R}$ depending on the content distributions. We separately discuss the cache-related SINR coverage probabilities for two tiers in this section, which is the theoretical basis for analyzing the final performance considering the association strategies.

Based on the content placement, $\Phi_{i,f}$ can be regarded as an independent non-HPPP with the density ${p_{i,f}}{\lambda _i}$. Therefore, the PDF of the distance between the typical UE and its nearest BS that contains the $f$-th file is given by
\begin{align}
P_{i,f}(r) = 2\pi{p_{i,f}}{\lambda _i}r\exp(-\pi{p_{i,f}}{\lambda _i}r^2), \ \ \ (r \geq 0).
\end{align}

In the following part, we first analyze the cache-related SINR coverage performance in the pico tier and then we consider the macro tier.

\subsection{SINR Coverage Analysis in The Second Tier}

If the typical UE is associated with the pico tier, the corresponding cache-related coverage probability can be derived with the aid of \emph{Laplace Transform of Interference}.
\subsubsection{Laplace Transform of Interference}
Since the path loss exponent of LOS links $\alpha_2$ is no less than 2 for practical scenarios, we divide the analysis on the Laplace transform of interference into two conditions ($\alpha_2>2$ and $\alpha_2=2$) in order to achieve closed-form expressions.
\begin{lemma}\label{lemma1}
\emph{When requesting the $f$-th content, under the condition $\alpha_2>2$, the Laplace transform of interference with the pre-decided SINR threshold $\tau$ in the pico tier is as follows}
\begin{align}\label{12}
\mathcal{L}_{2}( {s,\tau } ) =& \exp \Big( { - \pi {\lambda _2}( {R_L^2 - p_{2,f}{r^2}} ) }\nonumber \\
&{- \frac{{{\pi ^2}{\lambda _2}}}{{2u_1}}\sum\limits_{k_1 = 1}^{u_1} {{W_{f}}\big( {\frac{{{x_{k_1}}d}}{\lambda },s,\tau } \big)} \sqrt {( {1 - x_{k_1}^2} )} } \Big),
\end{align}
\emph{where}
\begin{align}\label{a1}
&{W_f}( {\omega ,s,\tau }) = p_{2,f}{S_2^0}\big( {\frac{{s{G_2}(\omega )\tau }}{{{N^p_2}{r^{{\alpha _2}}}}}} \big){r^2} \nonumber \\
&+\left( {1 - {p_{2,f}}} \right){\Delta _2}\left( {\frac{{s{G_2}\left( \omega  \right)\tau }}{{N_2^p}}} \right)- {S_2^0}\big( {\frac{{s{G_2}(\omega )\tau }}{{{N^p_2}R_L^{{\alpha _2}}}}} \big)R_L^2,
\end{align}
\emph{$x_{k_1}=cos(\frac{{2{k_1}- 1}}{{2{u_1}}}\pi ),k_1=1,2,...,u_1,$ are Gauss-Chebyshev nodes over $[-1,1]$, and $u_1$ is a trade-off parameter between the accuracy and complexity~\cite{7812773,7445146}. When $u_1\rightarrow \infty$, the equality is established. ${S_i^k}( z ){ = _2}{F_1}\big( {k - \frac{2}{{{\alpha _i}}},k + {N^p_i};k + 1 - \frac{2}{{{\alpha _i}}}; - z} \big)$ and ${}_2{F_1}( . )$ denotes Gauss hypergeometric function. ${\Delta _i}(z) = \Gamma \left( {1 - \frac{2}{{{\alpha _i}}}} \right)\Gamma \left( {N_i^p + \frac{2}{{{\alpha _i}}}} \right)\Gamma {\left( {N_i^p} \right)^{ - 1}}{z^{\frac{2}{{{\alpha _i}}}}}$ and $\Gamma(.)$ is the gamma function.}

\emph{Numerous actual channel measures~\cite{deng201528,rappaport201238} have indicated that the LOS path loss exponent is $2$ for various carrier frequencies, e.g. $28$ GHz, $38$ GHz and $73$ GHz. Under such condition $(\alpha_2=2)$, the equation~\eqref{a1} is changed to}
\begin{align}\label{a1_2}
&{W_f}( {\omega ,s,\tau } ) \nonumber \\
=& \frac{{s{G_2}( \omega  )\tau }}{{{N^p_2}}}\Big( {{F_y}\big( {\frac{{s{G_2}( \omega  )\tau }}{{{N^p_2}R_L^2}}} \big) - p_{2,f}{F_y}\big( {\frac{{s{G_2}( \omega  )\tau }}{{{N^p_2}{r^2}}}} \big)} \Big),
\end{align}
\emph{where}
\begin{align}
{F_y}( y ) =& N_2^p\ln \big( {1 + \frac{1}{y}} \big) - \frac{1}{{y{{( {1 + y} )}^{N_2^p - 1}}}}\nonumber \\
 &- \sum\limits_{m = 1}^{N_2^p - 1} {\frac{{N_2^p}}{{{{( {1 + y} )}^{N_2^p - m}}( {N_2^p - m} )}}} .
\end{align}
\begin{IEEEproof}
See Appendix A.
\end{IEEEproof}
\end{lemma}
\begin{remark}\label{remark1}
 The analytical expressions in \emph{\textbf{Lemma~\ref{lemma1}}} show that $\mathcal{L}_{2}( {s,\tau } )$ is independent of the transmit power $P_2$ and the intercept $C_2$.
\end{remark}
\begin{remark}\label{remark2}
With the aid of \eqref{12} and \eqref{a1_2}, we conclude that $\mathcal{L}_{2}( {s,\tau } )$ is a monotonic increasing function with $p_{2,f}$.
\end{remark}
\subsubsection{Cache-Related Coverage Probability}
Considering the biased received power, we define the cache-related coverage probability $\dot{P}_{{\Upsilon _{2,f}}}( {{\tau}} )$, when requesting the $f$-th file from the second tier, as follows
\begin{align}\label{coverage1}
\dot{P}_{{\Upsilon _{2,f}}}( {{\tau}} ) = \mathbb{P}\big[ {\Upsilon_{2,f} > {\tau}} \big].
\end{align}
where $\mathbb{P}(.)$ represents the probability function. With the aid of \textbf{Lemma~\ref{lemma1}}, the closed-form coverage probability of pico tier is calculated as below.
\begin{theorem}\label{theorem1}
\emph{When the typical UE requests the $f$-th ranked content from the pico tier, the cache-related SINR coverage probability ${\dot{P}_{{\Upsilon _{2,f}}}}$ in this dense mmWave network can be expressed as follows}
\begin{align}
\dot{P}_{{\Upsilon _{2,f}}}( {{\tau}} ) \approx &\frac{{\pi {R_L}}}{{2u_2}}\sum\limits_{n = 1}^{N_2^p} {{{( { - 1} )}^{n + 1}} {{N_2^p} \choose n}}\nonumber \\
&\times{\sum\limits_{k _2= 1}^{u_2} {{\dot{F_D}}\Big( {\frac{{( {{x_{k_2}} + 1} ){R_L}}}{2},\tau } \Big)} \sqrt {( {1 - x_{k_2}^2} )} } ,
\end{align}
\emph{where}
\begin{align}\label{17}
{\dot{F_D}}( {r,\tau } ) = \mathcal{L}_2\big( {\frac{{n\eta_L {r^{{\alpha _2}}}}}{{{G_0}}},\tau } \big)\exp \big( { - \frac{{n{\eta _L}{r^{{\alpha _2}}}\tau \sigma _2^2}}{{{P_2}{C_2}{N_2}{G_0}}}} \big){P_{2,f}}( r ),
\end{align}
\emph{and $\eta _L=N^p_2(N^p_2!)^{-1/N^p_2}$.}
\begin{IEEEproof}
Note that in equation \eqref{coverage1}, $|h_2|^2$ is a normalized Gamma random variable with parameter $N^p_2$. With the aid of \emph{\textbf{Lemma~\ref{lemma1}}} and the tight upper bound equation in Appendix~A from~\cite{6932503} $(\mathbb{P}\left[ {{{\left| {{h_2}} \right|}^2} < \gamma } \right] < {\left( {1 - \exp \left( { - {\eta _L}\gamma } \right)} \right)^{N_2^P}}, \gamma > 0)$, the cache-related SINR coverage probability is given by
\begin{align}
 \dot{ P}_{{\Upsilon _{2,f}}}( {{\tau}} )\approx &\sum\limits_{n = 1}^{{N^p_2}} {{{( { - 1} )}^{n + 1}}{N^p_2 \choose n}\int_0^{{R_L}} {\mathcal{L}_2\big( {\frac{{{n\eta _L}{r^{{\alpha _2}}}}}{{{G_0}}}} ,\tau \big)} }\nonumber\\
 &\times{\exp \big( { - \frac{{n{\eta _L}{\tau}\sigma _2^2{r^{{\alpha _2}}}}}{{{P_2}{C_2}{N_2}{G_0}}}} \big)} {\rm{}}{P_{2,f}}( r )dr.
\end{align}
By applying Gauss-Chebyshev Quadrature, we obtain \emph{\textbf{Theorem~\ref{theorem1}}}.
\end{IEEEproof}
\end{theorem}
\begin{remark}
When $p_{2,f}=1$, $\dot{P}_{{\Upsilon _{2,f}}}( {{\tau}} )$ in \emph{\textbf{Theorem~\ref{theorem1}}} represents the coverage probability of this dense mmWave network without caching ability.
\end{remark}
\begin{remark}
If the pico tier is assumed to be a noise-limited system, the signal-to-noise-ratio (SNR) coverage probability can be effortlessly deriving from \emph{\textbf{Theorem~\ref{theorem1}}} by deleting the interference part $\mathcal{L}_2(.)$ in equation~\eqref{17}.
\end{remark}

Due to the long communicating distance and high path loss, traditional cellular networks with mmWave are noise-limited~\cite{7493676}. However, recent articles~\cite{andrews2011tractable,8016632} have shown that with a high BS density, such systems become interference-limited. Under this condition, we present the first assumption below and the corroboration is provided in Section~\ref{numerical}.
\begin{assumption}\label{assumption1}
\emph{The dense mmWave network in the pico tier is assumed to be an interference-limited system, $\sigma_2^2=0$.}
\end{assumption}
\begin{corollary}\label{corollary1}
\emph{Under \textbf{Assumption~\ref{assumption1}}, the corresponding cache-related SINR coverage probability ${P_{{\Upsilon _{2,f}}}}$ in the dense mmWave network can be simplified as follows}
\begin{align}
{P_{{\Upsilon _{2,f}}}}( \tau  ) \approx &\frac{{\pi {R_L}}}{{2u_2}}\sum\limits_{n = 1}^{N_2^p} {{{( { - 1} )}^{n + 1}}{N^p_2 \choose n}}\nonumber\\
&\times{\sum\limits_{k_2 = 1}^{u_2} {{{ F}_D}\Big( {\frac{{( {{x_{k_2}} + 1} ){R_L}}}{2},\tau } \Big)\sqrt {( {1 - x_{k_2}^2} )} } } ,
\end{align}
\emph{where}
\begin{align}
{ F_D}( r,\tau ) = \mathcal{L}_2( {\frac{{{n\eta _L}{r^{{\alpha _2}}}}}{{{G_0}}}},\tau ){P_{2,f}}( r ).
\end{align}
\begin{IEEEproof}
By deleting the part $\big(\exp \big( { - \frac{{n{\eta _L}{\tau}\sigma _2^2{r^{{\alpha _2}}}}}{{{P_2}{C_2}{N_2}{G_0}}}} \big)\big)$ in \emph{\textbf{Theorem~\ref{theorem1}}}, which represents the thermal noise effect on the coverage performance, we obtain the equations for this corollary.
\end{IEEEproof}
\end{corollary}
\begin{remark}\label{remark4}
Based on \emph{\textbf{Remark~\ref{remark1}}}, we conclude that ${P_{{\Upsilon _{2,f}}}}( \tau  )$ is independent of $P_2$ and $C_2$. Moreover, note that \emph{\textbf{Corollary~\ref{corollary1}}} has a negligible difference with the exact simulations (as illustrated in Section~\ref{numerical}). We utilize such simplified expression as a replacement of the exact one in the rest of this paper.
\end{remark}

Since the association probability of Max-Rate is deduced on the basis of the derivative of the coverage probability~\cite{7493676}, we deduce the derivative of $W_f(.)$ described in \textbf{Lemma~\ref{lemma1}}, based on which the PDF of the coverage probability can be figured out.
\begin{lemma}\label{lemma2}
\emph{As $s$ in $W_f(\omega,s,\tau)$ is the transform variable of $r$ in our calculation, we introduce $s= {{{n\eta _L}{r^{{\alpha _2}}}}}/{{{G_0}}}$ into the equations to make the notation straightforward. When $\alpha_2>2$, the derivative of $W_f(\omega,r,\tau)$ is given by}
\begin{align}
&{w_f}(\omega ,r,\tau ) = \nonumber \\
&\frac{{2N_2^pZ\left( \omega  \right)}}{{\left( {{\alpha _2} - 2} \right)}}\left( {S_2^1\left( {Z(\omega )\tau } \right){r^2} - {p_{2,f}}S_2^1\left( {\frac{{Z(\omega )\tau {r^{{\alpha _2}}}}}{{R_L^{{\alpha _2}}}}} \right)\frac{{{r^{{\alpha _2}}}}}{{R_L^{{\alpha _2} - 2}}}} \right) \nonumber \\
&+\left( {1 - {p_{2,f}}} \right){\rm{Z}}\left( \omega  \right){\Lambda _2}\left( {{\rm{Z}}\left( \omega  \right)\tau } \right){r^{{2}}}
\end{align}
\emph{and when $\alpha_2=2$, such derivative can be expressed as}
\begin{align}
{w_f}(\omega ,r,\tau ) & =Z(\omega ){r^2}\nonumber \\
&\times\left( {{f_y}\left( {\frac{{Z(\omega )\tau {r^2}}}{{R_L^2}},r} \right) - {p_{2,f}}{f_y}\left( {Z(\omega )\tau ,{R_L}} \right)} \right),
\end{align}
\emph{where}
\begin{align}
{f_y}(y,z) =&{F_y}(y) + {Z}(\omega )\tau \frac{{{z^2}}}{{R_L^2}}\Big( {\frac{{N_2^py + 1}}{{{y^2}{{(1 + y)}^{N_2^p}}}}} \nonumber \\
&{- \frac{{N_2^p}}{{y(1 + y)}} + \sum\limits_{m = 1}^{N_2^p - 1} {\frac{{N_2^p}}{{{{(1 + y)}^{N_2^p - m + 1}}}}} } \Big),
\end{align}
\emph{${\Lambda _i}\left( z \right) = \frac{2}{{{\alpha _i}}}\Gamma \left( {1 - \frac{2}{{{\alpha _i}}}} \right)\Gamma \left( {N_i^p + \frac{2}{{{\alpha _i}}}} \right)\Gamma {\left( {N_i^p} \right)^{ - 1}}{z^{\frac{2}{{{\alpha _i}}} - 1}}$ and ${Z}( \omega  ) = \frac{{n{\eta _L}{G_2}( \omega  )}}{{{G_0}{N^p_2}}}$.}
\begin{IEEEproof}
With the fact that $\frac{d}{{dz}}S_i^0(z) =  \frac{{2N_i^P S_i^1(z)}}{{( {{\alpha _i}-2 } )}}$ and $\frac{d}{{dz}}\Delta_i(z)=\Lambda (z)$, we deduce the derivative of equations~\eqref{a1} and~\eqref{a1_2} under two conditions $(\alpha_2 >2$ and $\alpha_2=2)$.
\end{IEEEproof}
\end{lemma}
\begin{corollary}\label{corollary2}
\emph{With the aid of \textbf{Lemma~\ref{lemma2}}, when the typical UE requires the $f$-th ranked content, the PDF of cache-related SINR coverage probability for the second tier ${p_{{\Upsilon _{2,f}}}}$ is as follows}
\begin{align}
{p_{{\Upsilon _{2,f}}}}( \tau  ) \approx &\frac{{\pi {R_L}}}{{2u_3}}\sum\limits_{n = 1}^{N_2^p} {{{( { - 1} )}^{n + 1}}{N^p_2 \choose n}}\nonumber \\
&\times{\sum\limits_{k_3 = 1}^{u_3} {{{ f}_D}\big( {\frac{{{R_L}( {{x_{k_3}} + 1} )}}{2},\tau } \big)\sqrt {( {1 - x_{k_3}^2} )} } } ,
\end{align}
\emph{where}
\begin{align}
{{ f}_D}( {r,\tau } ) = \frac{{{\pi ^2}{\lambda _2}{{ F}_D}( {r,\tau } )}}{{2u_2}}\sum\limits_{k_2 = 1}^{u_2} {{w_f}\big( {\frac{{{x_{k_2}}d}}{\lambda },r,\tau } \big)\sqrt {( {1 - x_{k_2}^2} )} } .
\end{align}
\begin{IEEEproof}
See Appendix B.
\end{IEEEproof}
\end{corollary}
\subsection{SINR Coverage Analysis in The First Tier}
In the macro tier, we utilize the Rayleigh fading channel for sub-6 GHz signals. The exact expression of cache-related SINR coverage probability can be achieved in this part.
\subsubsection{Cache-Related Coverage Probability}
As discussed in the previous part, we keep the SINR threshold $\tau$ in the sub-6 GHz tier. With the similar analysis in the pico tier, the cache-related coverage probability in the second tier for requiring the $f$-th ranked content can be expressed as follows
\begin{align}\label{coverage}
\dot{P}_{{\Upsilon _{1,f}}}( {{\tau}} ) =\mathbb{P}\left[ {\Upsilon_{1,f} > {\tau}} \right].
\end{align}

Due to the Rayleigh fading channel, it is effortless to derive the Laplace transform of interference for the first tier. Therefore, we directly provide the cache-related coverage probability in the following paragraph.
\begin{theorem}\label{theorem2}
\emph{When the typical UE requests the $f$-th ranked file from the macro tier, the exact cache-related SINR coverage probability ${\dot{P}_{{\Upsilon _{1,f}}}}$ is given by}
\begin{align}
&{\dot{P}_{{\Upsilon _{1,f}}}}( {{\tau}} )  = \int_0^\infty  {\exp \big( { - \frac{{\tau \sigma _1^2{r^{{\alpha _1}}}}}{{{P_1}{C_1}{N_1}G_0}} { - \pi {\lambda _1}{r^2}}}}\nonumber\\
&{{{\times\left( {{p_{1,f}}(S_1^0(\tau ) - 1) + (1 - {p_{1,f}}){\Delta _1}(\tau )} \right)}} \big){P_{1,f}}( r )dr} .
\end{align}
\begin{IEEEproof}
As shown in equation \eqref{coverage}, ${\dot{P}_{{\Upsilon _{1,f}}}}( \tau  ) = \mathbb{P}\left[ {|{h_1}{|^2} > \frac{{\tau ( {{I_{1,f}}+I_{1,f^\prime} + \sigma _1^2} ){r^{{\alpha _1}}}}}{{C_1{N_1}{G_0}}}|r = ||{x_0}||} \right]$ where $|h_1|^2 \sim \exp (1)$ due to Rayleigh fading assumption. Thus $\dot{P}_{{\Upsilon _{1,f}}}( {{\tau}} )$ can be expressed as $\dot{P}_{{\Upsilon _{1,f}}}( {{\tau}} ) = \mathbb{E}\left[ {\exp ( { - \frac{{\tau ( {{I_{1,f}} +I_{1,f^\prime} + \sigma _1^2} ){r^{{\alpha _1}}}}}{{C_1{N_1}{G_0}}}} )|r = ||{x_0}||} \right]$. Using the same method of \emph{\textbf{Lemma~\ref{lemma1}}}, the coverage probability can be figured out as shown above.
\end{IEEEproof}
\end{theorem}

\emph{Special Case 1:} We assume $\alpha_2=4$, as it is valid for most sub-6 GHz networks~\cite{7446343}.
\begin{corollary}\label{corollary3}
\emph{Under \emph{Special Case 1}, the closed-form cache-related coverage probability in the first tier can be expressed as follows}
\begin{align}
{\tilde{P}_{{\Upsilon _{1,f}}}}( {{\tau}} ) = \frac{1}{2}\pi p_{1,f}{\lambda _1}\sqrt {\frac{\pi }{{B( {{\tau}} )}}} \exp \big( {\frac{{{C^2}( {{\tau}} )}}{{4B( {{\tau}} )}}} \big)\erfc \big( {\frac{{C( {{\tau}} )}}{{2\sqrt {B( {{\tau}} )} }}} \big),
\end{align}
\emph{where $B( {{\tau}} ) = \frac{{{\tau}\sigma _1^2}}{{P_1{C_1}{N_1}{G_0}}}$, $C( \tau  ) = \pi {\lambda _1}( {{p_{1,f}}{S_1^0}( \tau  ) +(1-p_{1,f})\Delta_1(\tau) } )$, and $\erfc (.)$ is the complementary error function.}
\begin{IEEEproof}
Note that $\alpha_1=4$ as mentioned in Special Case 1, the equation of \emph{\textbf{Theorem~\ref{theorem2}}} can be simplified as ${\tilde{P}_{{\Upsilon _{1,f}}}}( {{\tau}} ) = 2\pi p_{1,f} {\lambda _1}\int_0^\infty  {\exp ( { - B( {{\tau}} ){r^4} - C( {{\tau}} ){r^2}} )} rdr$. Deploying (3.462-1) in~\cite{jeffrey2007table}, we have a closed-form expression ${\tilde{P}_{{\Upsilon _{1,f}}}}( {{\tau}} ) = \pi p_{1,f}{\lambda _1}{( {2B( {{\tau}} )} )^{ - \frac{1}{2}}}\exp ( {\frac{{{C^2}( {{\tau}} )}}{{8B( {{\tau}} )}}} ){D_{ - 1}}( {\frac{{C( {{\tau}} )}}{{\sqrt {2B( {{\tau}} )} }}} )$, where ${D_{n}}( z )$ is the Parabolic cylinder functions and ${D_{ - 1}}( z ) = \exp ( {\frac{{{z^2}}}{4}} )\sqrt {\frac{\pi }{2}} \erfc( {\frac{z}{{\sqrt 2 }}} )$.
\end{IEEEproof}
\end{corollary}
\begin{assumption}\label{assumption2}
Since in various articles~\cite{7446343,7493676,liu2017caching}, the noise can be ignored in the traditional cellular networks with sub-6 GHz, we assume only the signal-to-interference-ratio (SIR) is considered in the first tier, namely, $\sigma _1^2=0$.
\end{assumption}
\begin{corollary}\label{corollary4}
\emph{Under \textbf{Assumption~\ref{assumption2}}, the cache-related coverage probability for the first tier in \textbf{Theorem~\ref{theorem2}} can be simplified as follows}
\begin{align}
{P_{{\Upsilon _{1,f}}}}(\tau ) = \frac{{{p_{1,f}}}}{{{p_{1,f}}S_1^0(\tau ) + (1 - {p_{1,f}}){\Delta _1}(\tau )}}.
\end{align}
\begin{IEEEproof}
By removing the noise part $\big(\exp(- \frac{{\tau \sigma _1^2{r^{{\alpha _1}}}}}{{P_1{C_1}{N_1}{G_0}}})\big)$ from \emph{\textbf{Theorem~\ref{theorem2}}}, the SIR coverage probability can be deduced with the fact that $\int_0^\infty  {r\exp ( - a{r^2})dr = } \frac{1}{{2a}},a > 0$.
\end{IEEEproof}
\end{corollary}
\begin{remark}\label{remark6}
${P_{{\Upsilon _{1,f}}}}(\tau )$ is a monotonic increasing function with $p_{1,f}$. Moreover, ${P_{{\Upsilon _{1,f}}}}(\tau )$ is independent of $\lambda_1$, $P_1$, $N_1$, and $C_1$. Due to the negligible difference with the simulation shown in Section~\ref{numerical}, we use this closed-form equation as a proxy of the exact expression in the remainder of this paper.
\end{remark}
With the aid of the closed-form expression in \textbf{Corollary~\ref{corollary4}}, we are able to figure out the closed-form derivative of cache-related coverage probability for the first tier effortlessly.
\begin{corollary}\label{corollary5}
\emph{Under \textbf{Assumption~\ref{assumption2}}, the PDF ${p_{{\Upsilon _{1,f}}}}(\tau )$ of cache-related coverage probability when requesting the $f$-th ranked file is shown as follows}
\begin{align}
{p_{{\Upsilon _{1,f}}}}(\tau ) = \frac{{2p_{1,f}^2S_1^1(\tau ) + \left( {{\alpha _1} - 2} \right)({p_{1,f}} - p_{1,f}^2){\Lambda _1}\left( \tau  \right)}}{{\left( {{\alpha _1} - 2} \right){{\left( {{p_{1,f}}S_1^0(\tau ) + (1 - {p_{1,f}}){\Delta _1}(\tau )} \right)}^2}}}.
\end{align}
\begin{IEEEproof}
As the PDF of coverage probability for the second tier is ${p_{{\Upsilon _{1,f}}}}( \tau  ) =  - \frac{{d{P_{{\Upsilon _{1,f}}}}( \tau  )}}{{d\tau }}$, we are able to calculate the expression based on \emph{\textbf{Corollary~\ref{corollary4}}}.
\end{IEEEproof}
\end{corollary}

\section{Success Probability and Area Spectral Efficiency Analysis}\label{Success_ASE}

From the perspective of customers, the success probability is an important parameter to appraise the quality of service. In our cache-enabled HetNet, the data rate at the typical UE exceeding the pre-decided rate threshold $R_{th}$ contributes to the success probability~\cite{liu2017caching}.

As discussed in the previous sections, we conclude that the considered system has two different processes in sending multimedia contents: 1) \emph{Association Mode}, when the requested $f$-th file obeys $(1\leq f\leq H_c)$, the typical UE chooses the suitable BS as the corresponding BS depending on two association strategies; and 2) \emph{Server Mode}, when the demanded $f$-th content only exists in the server due to limited storage capacity at BSs, which means $(H_c< f\leq N_c)$, the typical UE requests such content from the server via the nearest macro BS. We detailedly discuss these two modes below.

\subsection{Association Mode}

In this mode, since two association strategies (Max-RP and Max-Rate) have different judgment standards to decide the corresponding BS, we study them separately.
\subsubsection{Maximum Received Power Scheme}
The Max-RP scheme has been utilized in numerous HetNets proposed in recent articles, for example, the traditional cache-enabled HetNets~\cite{liu2017caching} and the hybrid HetNets with mmWave~\cite{7493676}. Under this scheme, the association procedure is fast and at a low cost due to ignoring the interference effects. We define the Max-RP association probability, when the typical UE connects to the $i$-th tier BS for the $f$-th file, as follows
\begin{align}
{\mathcal{A}^P_{i,f}} = \mathbb{P}\left[ \bar P_{i,f} > \bar P_{j,f} \right],
\end{align}
where $j\neq i$ and $j\in [1,2]$.

Note that the path loss law for the pico tier $L_2(r)$ has a step character. With the aid of similar proof in~\cite{jo2012heterogeneous}, the PDF $f_{i,f}^P( r )$ of the distance $r$ between the typical UE and its corresponding $i$-th tier BS with containing $f$-th ranked file under Max-RP scheme is given by
\begin{align}\label{distance_distribution}
&f_{1,f}^P\left( r \right) =\nonumber \\
& 2\pi {p_{1,f}}{\lambda _1}r\exp \left( { - \pi \sum\limits_{j = 1}^2 {{p_{j,f}}{\lambda _j}P{{_{j,1}^r}^{\frac{2}{{{\alpha _2}}}}}{r^{\frac{2}{{\alpha _{j,1}^r}}}}} } \right)\mathbf{U}\left( {{R_r} - r} \right) \nonumber \\
&+ 2\pi {p_{1,f}}{\lambda _1}r\exp \left( { - \pi {p_{2,f}}{\lambda _2}R_L^2 - \pi {p_{1,f}}{\lambda _1}{r^2}} \right)\mathbf{U}\left( {r - {R_r}} \right),\\
&f_{2,f}^P\left( r \right) = \nonumber \\
&2\pi {p_{2,f}}{\lambda _2}r\exp \left( { - \pi \sum\limits_{j = 1}^2 {{p_{j,f}}{\lambda _j}P{{_{j,2}^r}^{\frac{2}{{{\alpha _1}}}}}{r^{\frac{2}{{\alpha _{j,2}^r}}}}} } \right)\mathbf{U}\left( {{R_L} - r} \right),
\end{align}
where ${{P}^r_{j,i}} = \frac{{b_j^PP_jC_jN_j}}{{b_i^PP_iC_iN_i}}$, ${{ \alpha }^r_{j,i}} = \frac{{{\alpha _j}}}{{{\alpha _i}}}$, and ${R_r} = {\left( {P_{1,2}^rR_L^{{\alpha _2}}} \right)^{\frac{1}{{{\alpha _1}}}}}$.

\subsubsection{Maximum Rate Scheme}
Since the path loss laws and bandwidth for two tiers are dissimilar, the received data rates are totally different even they have the same average received power~\cite{7493676}. Compared with Max-RP, Max-Rate is able to provide higher data rate, but the extra knowledge of channel state information is indispensable. For Max-Rate, the association probability of the typical UE being associated with the $i$-th tier BS for requesting the $f$-th file is defined as
\begin{align}
{\mathcal{A}^R_{i,f}} = \mathbb{P}\left[ R_{i,f} > R_{j,f} \right].
\end{align}

Instead of analyzing the relationship between the PDF of the corresponding distance as discussed in Max-RP, we are able to directly derive the PDF of considered coverage probability with the SINR threshold $\tau$.
\begin{lemma}\label{lemma3}
\emph{When requesting the $f$-th ranked content, the PDF of the cache-related coverage probability for the $i$-th tier under the Max-Rate strategy is shown as follows}
\begin{align}
f_{i,f}^R( \tau  ) = {p_{{\Upsilon _{i,f}}}}( \tau  )\Big( {1 - {P_{{\Upsilon _{j,f}}}}\big( {{{( {1 + \tau } )}^{\frac{{{b_i^BB_i}}}{{{b_j^BB_j}}}}} - 1} \big)} \Big).
\end{align}
\begin{IEEEproof}
See Appendix C.
\end{IEEEproof}
\end{lemma}
\begin{remark}\label{remark7}
Since ${P_{{\Upsilon _{j,f}}}}(\infty)=0$, if $(b_i^BB_i\gg b_j^BB_j)$, the PDF of this coverage probability is same as ${p_{{\Upsilon _{i,f}}}}( \tau  )$, which means the typical UE is associated with the $i$-th tier BS invariably. As a consequence, if the bandwidth of one tier is far more than the other, the typical UE always connects to the tier with large bandwidth.
\end{remark}

\noindent \textbf{Average Load Approximation}: When all UEs are associated with the HetNet, the average number of UEs served by the $i$-th tier BSs $N_i^{load}$ can be approximated by\footnote{This approximation is valid for sub-6 GHz scenarios~\cite{6497002} and mmWave scenarios~\cite{7110547}. Due to the low probability of requesting the content from the core server, we ignore the corresponding load in the server mode for simplifying the analysis.}~\cite{7493676} $\bar N_i^{load} \approx 1 + \frac{{1.28{\lambda _u}{\mathcal{A}_i}}}{{{\lambda _i}}},$ where $\mathcal{A}_i^P = \sum\limits_{f = 1}^{{H_c}} {{P_f}} \int_0^\infty  {f_{i,f}^P\left( r \right)} dr$, $\mathcal{A}_i^R = \sum\limits_{f = 1}^{{H_c}} {{P_f}} \int_{ - \infty }^{ + \infty } {f_{i,f}^R\left( \tau  \right)} d\tau $, and $\mathcal{A}_i\in\{\mathcal{A}^P_i,\mathcal{A}^R_i\}$.
\begin{remark}\label{remark8}
Since $\mathcal{A}_i$ is a monotonic increasing function with the corresponding bias factor $(b_i^P$ or $b^B_i)$, the small value of such bias factor is able to offload the data traffic in the $i$-th tier. As a result, by adjusting the bias factors $b_i^P$ in the Max-RP scheme and $b_i^B$ in the Max-Rate scheme, we are able to control the average number of UEs for each tier, thereby balancing the load of the proposed HetNet.
\end{remark}

\subsection{Server Mode}

We present the backhaul capacity in order to compare our system with the traditional HetNets~\cite{liu2017caching} in which the macro BSs have no caching ability. The comparison is illustrated in Section~\ref{numerical}. In the server mode, the backhaul capacity restricts the performance of our system. More specifically, if the required rate $R_{th}$ exceeds the backhaul capacity $C_{bh}$, no content can be sent successfully due to low system rate. On the other hand, if $C_{bh}$ is larger than $R_{th}$, the success probability under this case is limited by the received data rate from the relay macro BS. As a result, we provide the success probability in this mode as follows.
\begin{lemma}\label{lemma4}
\emph{The success probability ${P_S}( {{R_{th}}} ) $ in the server mode is given by}
\begin{align}
{P_S}({R_{th}}) =& \mathbf{U}({C_{bh}} - {R_{th}}) \mathbf{U}({N_{c}} - {H_{c}} - 1)\nonumber \\
&\times\sum _{f = {H_c} + 1}^{{N_c}}{P_f}{P_{{\Upsilon _{1}}}}({2^{\frac{{{R_{th}}}}{{{B_1}}}}} - 1),
\end{align}
\emph{where ${P_{{\Upsilon _{1}}}}(.)$ is the coverage probability of the first tier without cache capacity, which equals ${P_{{\Upsilon _{1,f}}}}(.)$ with the condition that $p_{1,f}=1$.}
\begin{IEEEproof}
As discussed above, if $C_{bh}<R_{th}$, ${P_S}( {{R_{th}}} )=0$, while if $C_{bh} \geq R_{th}$, the success probability for requesting the $f$-th file is decided by the wireless capacity of the macro tier, which is ${P_f}{P_{{\Upsilon _{1}}}}( {{2^{\frac{{{R_{th}}}}{{{B_1}}}}} - 1} )$. Then, we consider the request probability for the files from $(H_c+1)$ to $N_c$ to obtain this lemma (note that when $H_c=N_c$, the typical UE is able to acquire all files from the HetNet, namely ${P_S}( {{R_{th}}} )=0$).
\end{IEEEproof}
\end{lemma}

\subsection{Success Probability}

As mentioned in the beginning of this section, we present the success probability to evaluate the performance of our system. This parameter can be calculated with the aid of the cache-related SINR coverage probability discussed in Section III. We first define the universal success probability as below
\begin{align}
{{\cal P}_s}({R_{th}}) =& \sum\limits_{f = 1}^{H_c}\sum\limits_{i = 1}^2 P_f{{\cal A}_{i,f}^\kappa \mathbb{P}\left[ {{B_i}{{\log }_2}\left(1 + \Upsilon_{i,f}\right) > {R_{th}}} \right]}  \nonumber \\
&+ {P_S}({R_{th}}),
\end{align}
where $\kappa\in \{P,R\}$. We calculate the success probability under two association strategies in the following part.
\subsubsection{Maximum Received Power Scheme}
As the system rate can be derived from the received SINR, we first deduce the SINR coverage probability for the $i$-th tier under Max-RP.
\begin{lemma}\label{lemma5}
\emph{When requiring the $f$-th ranked content, the SINR coverage probability for the $i$-th tier ${\Theta _{i,f}}( \tau  )$ under Max-RP scheme is shown as}
\begin{align}
{\Theta _{1,f}}( \tau  ) =&\int_0^\infty  {\exp \big( { - \pi {\lambda _1}{r^2}\big( {{p_{1,f}}(S_1^0(\tau ) - 1) }}}\nonumber \\
&{{{+ (1 - {p_{1,f}}){\Delta _1}(\tau )} \big)} \big)} f_{1,f}^P(r)dr,
\end{align}
\begin{align}
{\Theta _{2,f}}( \tau  ) \approx &\frac{{\pi {R_L}}}{{2u_2}}\sum\limits_{n = 1}^{{N^p_2}} {{{( { - 1} )}^{n + 1}}{ N^p_2 \choose n }}\nonumber \\
&\times{\sum\limits_{k_2 = 1}^{u_2} {{F_R}\Big( {\frac{{( {{x_{k_2}} + 1} ){R_L}}}{2},\tau } \Big)} \sqrt {( {1 - x_{k_2}^2} )} } ,
\end{align}
\emph{where}
\begin{align}
{F_R}( {r,\tau } ) = \mathcal{L}_2\left( {\frac{{{n\eta _L}{r^{{\alpha _2}}}}}{{{G_0}}},\tau } \right)f_{2,f}^P( r ).
\end{align}
\begin{IEEEproof}
 Since the probability of distance under Max-RP scheme is changed into equation~\eqref{distance_distribution}, $P_{i,f}(r)$ is replaced by $f_{i,f}^R( r )$ in \emph{\textbf{Corollary~\ref{corollary1}}} and \emph{\textbf{Theorem~\ref{theorem2}}} for the second and first tiers, respectively. Moreover, with the aid of \emph{\textbf{Assumption 1}} and \emph{\textbf{Assumption 2}}, we also ignore the noise effect. Therefore, these probabilities are calculated as above.
\end{IEEEproof}
\end{lemma}
\begin{remark}
Since $f_{i,f}^R( . )$ has a negative correlation with ${P^r_{j,i}}$ in equation \eqref{distance_distribution}, ${\Theta _{i,f}}(.)$ is a monotonic increasing function with the product of $b_i^P$, $P_i$, and $C_i$.
\end{remark}

Then, based on the cache-related coverage probability under the Max-RP scheme, we present the corresponding success probability in the following part.

\begin{theorem}\label{theorem3}
\emph{With the aid of \textbf{Lemma 4} and \textbf{Lemma 5}, the success probability for Max-RP scheme can be expressed as follows}
\begin{align}
{\mathcal{P}_P}( {{R_{th}}} ) \approx \sum\limits_{f = 1}^{{H_c}} {\sum\limits_{i = 1}^2 {{P_f}} {\Theta _{i,f}}( {{2^{\frac{{{R_{th}}}}{{{B_i}}}}} - 1} )}  + {P_S}( {{R_{th}}} ).
\end{align}
\begin{IEEEproof}
When the required rate is $R_{th}$, the SINR threshold is $({{2^{\frac{{{R_{th}}}}{{{B_i}}}}} - 1} )$, so the success probability for requesting the $f$-th file from the first $H_c$ contents is $ {P_f} {\Theta _{i,f}}( {{2^{\frac{{{R_{th}}}}{{{B_i}}}}} - 1} ) $. Considering the whole multimedia contents, the success probability will be calculated as above.
\end{IEEEproof}
\end{theorem}
\subsubsection{Maximum Rate Scheme}
In contrast to the discussion in Max-RP scheme, we have already figured out PDF of the coverage probability for the $i$-th tier in \textbf{Lemma 3}. The success probability can be calculated based on this result.
\begin{theorem}
\emph{With the aid of \textbf{Lemma~\ref{lemma3}} and \textbf{Lemma~\ref{lemma4}}, the success probability in the Max-Rate scheme is given by}
\begin{align}
{\mathcal{P}_R}( {{R_{th}}} ) \approx \sum\limits_{f = 1}^{{H_c}} {\sum\limits_{i = 1}^2 {{P_f}\int_{{2^{\frac{{{R_{th}}}}{{{B_i}}}}} - 1}^\infty  {f_{i,f}^R( \tau  )d\tau } } }  + {P_S}( {{R_{th}}} ).
\end{align}
\begin{IEEEproof}
Note that the received SINR should be larger than the SINR threshold $({{2^{\frac{{{R_{th}}}}{{{B_i}}}}} - 1} )$. For the Max-Rate scheme, the success probability for requesting the $f$-th file is changed to ${P_f}\int_{{2^{\frac{{{R_{th}}}}{{{B_i}}}}} - 1}^\infty  {f_{i,f}^R( \tau  )d\tau }$. After considering both the association mode and the server mode, universal success probability can be deduced in this theorem.
\end{IEEEproof}
\end{theorem}
\begin{remark}\label{remark9}
Combining \emph{\textbf{Remark~\ref{remark4}}} and \emph{\textbf{Remark~\ref{remark6}}}, we conclude that the success probability under Max-Rate is independent of $\lambda_1$, $N_1$, $C_i$ and $P_i$.
\end{remark}

\subsection{Area Spectral Efficiency}

The ASE is the average data rate transmitted in unit bandwidth and unit area, which can be represented in the unit of bps/Hz/m$^2$. Assuming that Gaussian Codebooks are utilized for all transmissions, we are able to define ASE with the aid of Shannon's Capacity Formula. It is expressed as follows $ASE = \lambda {\log _2}\left( {1 + \tau } \right){P_{\tau}}$, where $P_{\tau}$ is the SINR coverage probability of the considered networks and $\lambda$ denotes the active BSs' density~\cite{7446343}.

\begin{proposition}
\emph{The $ASE$s for two user association strategies share the same equation as below}
\begin{align}\label{ASE}
ASE =& \sum\limits_{f = 1}^{{H_c}} {\sum\limits_{i = 1}^2 {\frac{{{P_f}{p_{i,f}}{\lambda _i}{R_{th}}}}{{{B_i}}}} } P_a^\kappa ({2^{\frac{{{R_{th}}}}{{{B_i}}}}} - 1)\nonumber \\
& + \frac{{{\lambda _1}{R_{th}}}}{{{B_1}}}{P_S}({R_{th}}),
\end{align}
\emph{where $P_a^P(\tau ) = {\Theta _{i,f}}(\tau )$ is the coverage probability when requiring the $f$-th file from the $i$-th tier under Max-RP and $P_a^R(\tau ) = \int_\tau ^\infty  {f_{i,f}^R(t)dt}$ is that under Max-Rate.}
\begin{IEEEproof}
When the requested $f$-th ranked file is located between the first and $H_c$-th popularity rank, the ASE with pre-decided SINR threshold $\tau$ for the $i$-th tier is $P_f{p_{i,f}}{\lambda _i}{\log _2}(1 + \tau )P_a^\kappa (\tau )$. On the other side, if $(f>H_c)$, the typical UE is associated with the macro tier to request the content from the server. Therefore, the ASE is $P_f{{\lambda _1}{{\log }_2}(1 + } \tau ){P_{{\Upsilon _{1}}}}(\tau )$. Then sum them up, we have the final equation above.
\end{IEEEproof}
\end{proposition}
\begin{remark}\label{remark10}
Note that an optimum value of $R_{th}$ exists for achieving the maximum ASE. The reason is that the large $R_{th}$ potentially increase the ASE as shown in \eqref{ASE}, but it also decreases the coverage probabilities of both tiers.
\end{remark}

\section{Numerical Results}\label{numerical}

\subsection{Network Settings and Simulations}

The general network settings are presented in Table~\ref{table1}~\cite{7493676,liu2017caching,6932503} and the reference distance for the intercept is one meter. Since we employ multi-input single-output (MISO) system at mmWave tier, the actual beamforming of ULA is imposed in Monte Carlo simulations. Additionally, the NLOS transmissions are provided in simulations as well for evaluating the effect of NLOS BSs. The path loss exponent is $\alpha_N=4$, the intercept is $C_N=C_2$ and the parameter $N_N$ for Nakagami fading channel in NLOS scenario is $2$~\cite{6932503}.
\begin{table*}[ht!]
\centering
\footnotesize
\caption{General Settings of the Network}
\label{table1}
\begin{tabular}{|l|l|}
\hline
   LOS ball range     & $R_L=200$ m\\ \hline
   Density of PPP     & $\lambda_1=1/ (250^2\pi)$; $\lambda_2=20/ (250^2\pi)$; $\lambda_u=30/ (250^2\pi)$ m$^{-2}$\\ \hline
   Bandwidth    & $B_1=20$ MHz; $B_2=1$ GHz\\ \hline
   Path loss law     & $\alpha_1=4$, $N^p_1=1$;  $\alpha_2=2$, $N^p_2=3$\\ \hline
   Number of antennas     & $N_1=1$; $N_2=20$\\ \hline
   Carrier frequency     & $f_{macro}=2$ GHz; $f_{pico}=28$ GHz\\ \hline
   Transmit Power at BSs         & $P_1=80$ dBm; $P_2=30$ dBm\\ \hline
   Transmit Power at the typical UE       & $P_0=30$ dBm\\ \hline
   Backhaul capacity      & $C_{bh}=50$ Mbps\\ \hline
   Caching capacity   & $M_1=M_2=80$, $H_c=90$, $N_c=100$\\ \hline
   Skew of the popularity distribution & $\delta=0.6$\\ \hline
   Bias factors & $b_i^P=b_i^B=1$\\ \hline
\end{tabular}
\end{table*}

\begin{figure*}[t!]
\centering
\subfigure[Cache-related coverage probability versus pre-decided SINR threshold, with $p_{i,f}=1$.]{\label{fig1} \includegraphics[width= 3.1in, height=2 in]{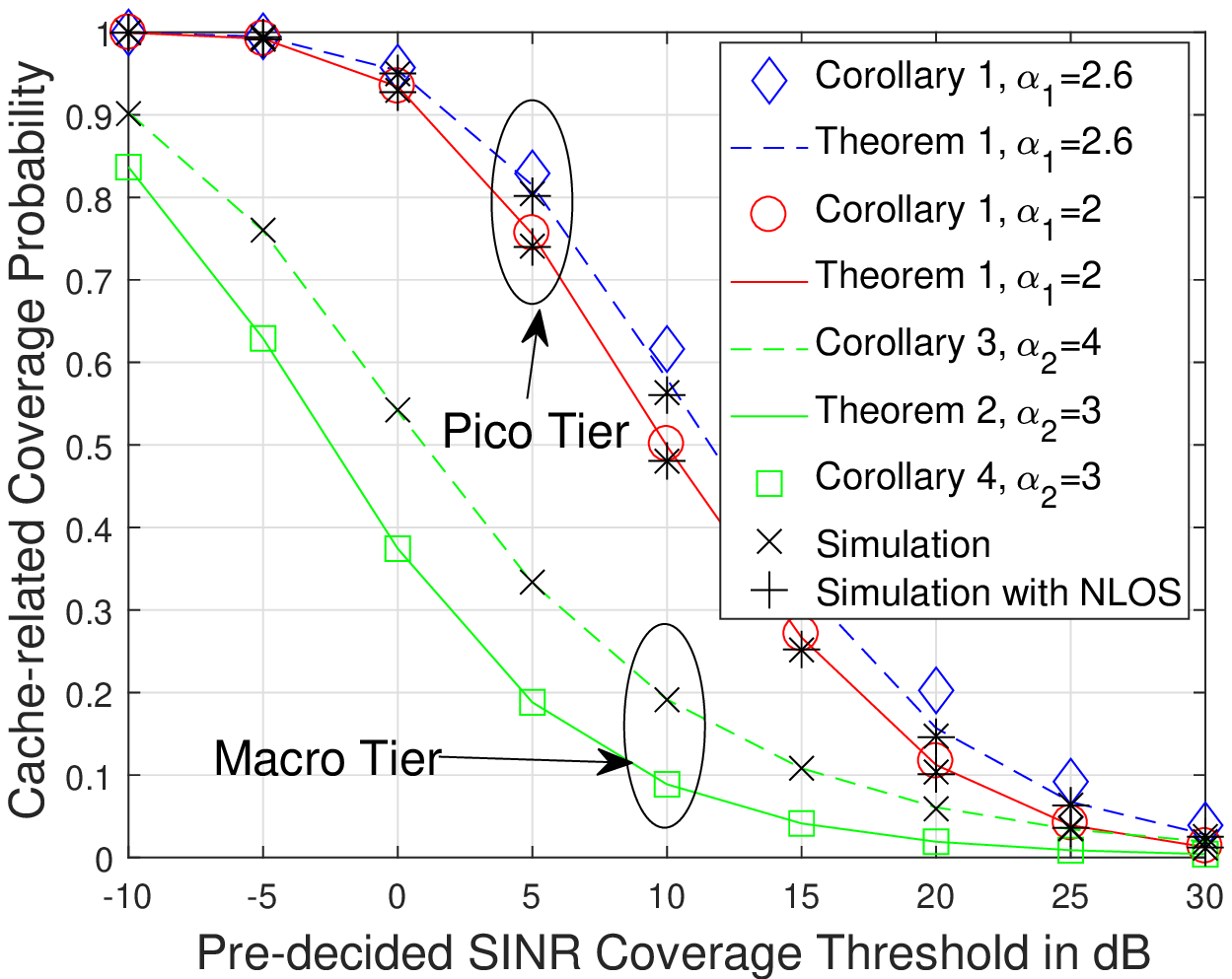}}
\subfigure[Success probability versus pre-decided rate threshold, with $M_2=10$, $B_1=200$ MHz, and $p_{i,f}=\textbf{U}(M_i-f)$.]{\label{fig2} \includegraphics[width= 3.1in, height=2  in]{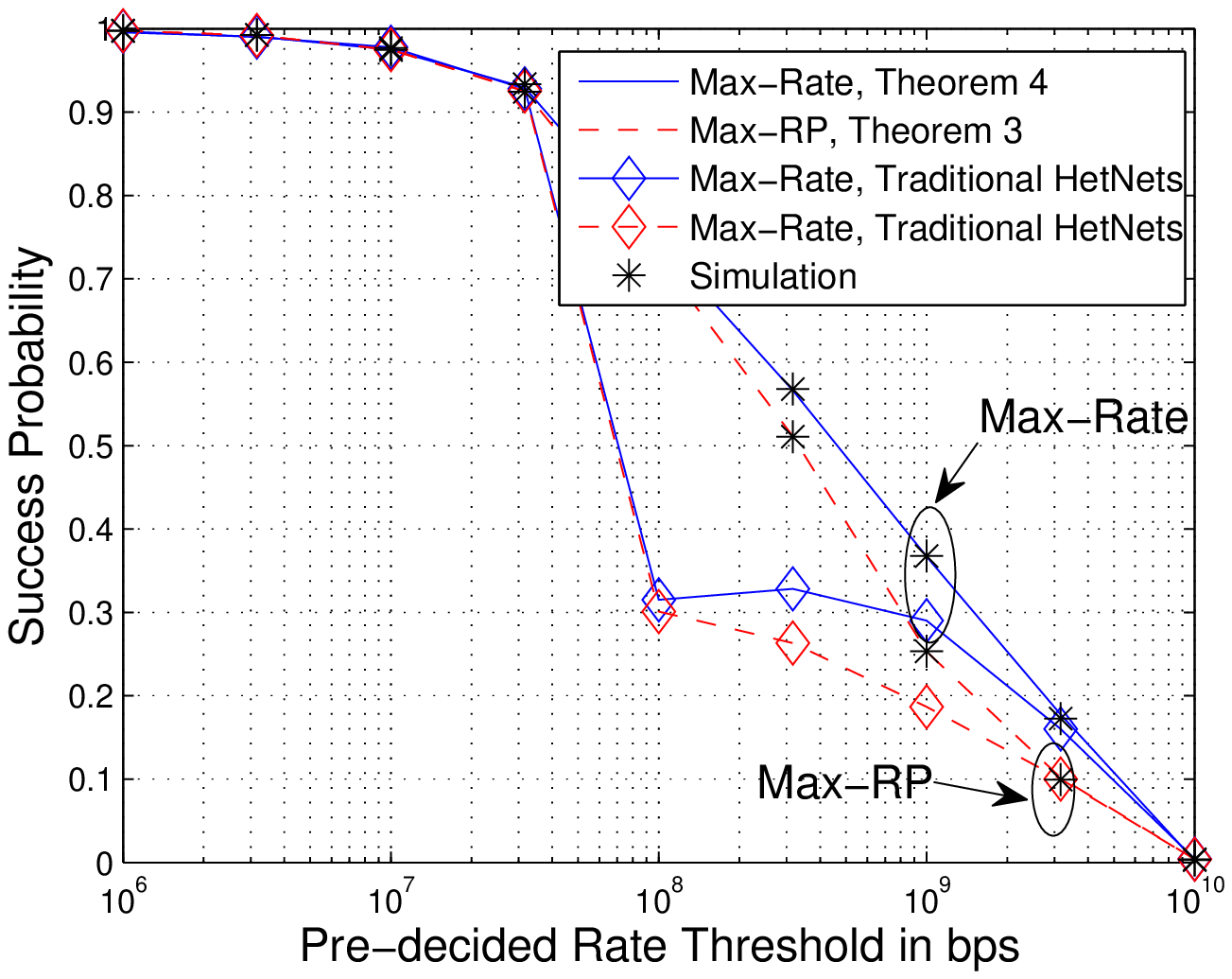}}
\caption{Simulation and Validation.}
\end{figure*}

In Fig.~\ref{fig1}, we validate the expressions of cache-related coverage probabilities for requesting the $f$-th file. For the second tier, the simulation results with NLOS BSs overlap those without NLOS transmissions, which means ignoring all NLOS signals does not impair the analytical accuracy. Our tight approximate equations for the second tier fit the simulation results with an insignificant difference, especially when the path loss exponent $\alpha_2=2$. With the increase of $\alpha_2$, the noise effect is slightly enhanced but our mmWave networks in the pico tier is still dominated by the interference. On the other hand, the theoretical expressions for the first tier match the numerical results perfectly, which means the macro tier with sub-6 GHz is an interference-limited system and the closed-form expression in \textbf{Corollary~\ref{corollary4}} is able to replace the exact equation in \textbf{Theorem~\ref{theorem2}} for simplifying the analysis. Moreover, the mmWave networks achieve a higher SINR coverage probability than the traditional sub-6 GHz networks under the same pre-decided coverage threshold, which indicates that large association probability to the pico tier benefits the universal throughput.

Comparing the analytical results of the success probability with the simulations in Fig.~\ref{fig2}, we note that they match each other ideally, thereby certifying the analysis. Max-Rate performs better than Max-RP in terms of the success probability when the pre-decided rate threshold $R_{th}$ is high. In the real world, we may use Max-RP as the best scheme for fetching the maximum success probability in the low $R_{th}$ region due to the easy operation with cheap system cost, while with the rise of $R_{th}$, the best strategy for user association in our system should be changed into Max-Rate although it needs extra system overheads. Additionally, it is obvious that our cache-enabled HetNets outperform the traditional HetNets, especially in the area $R_{th}\geq C_{bh}$.

\subsection{Impact of Noise and Antenna Scales}

\begin{figure*}[t!]
\centering
 \subfigure[Cache-related coverage probability versus pre-decided SINR threshold in the second tier, with $N_2=32$ and $p_{2,f}=1$.]{\label{fig3} \includegraphics[width= 3.1in, height=2 in]{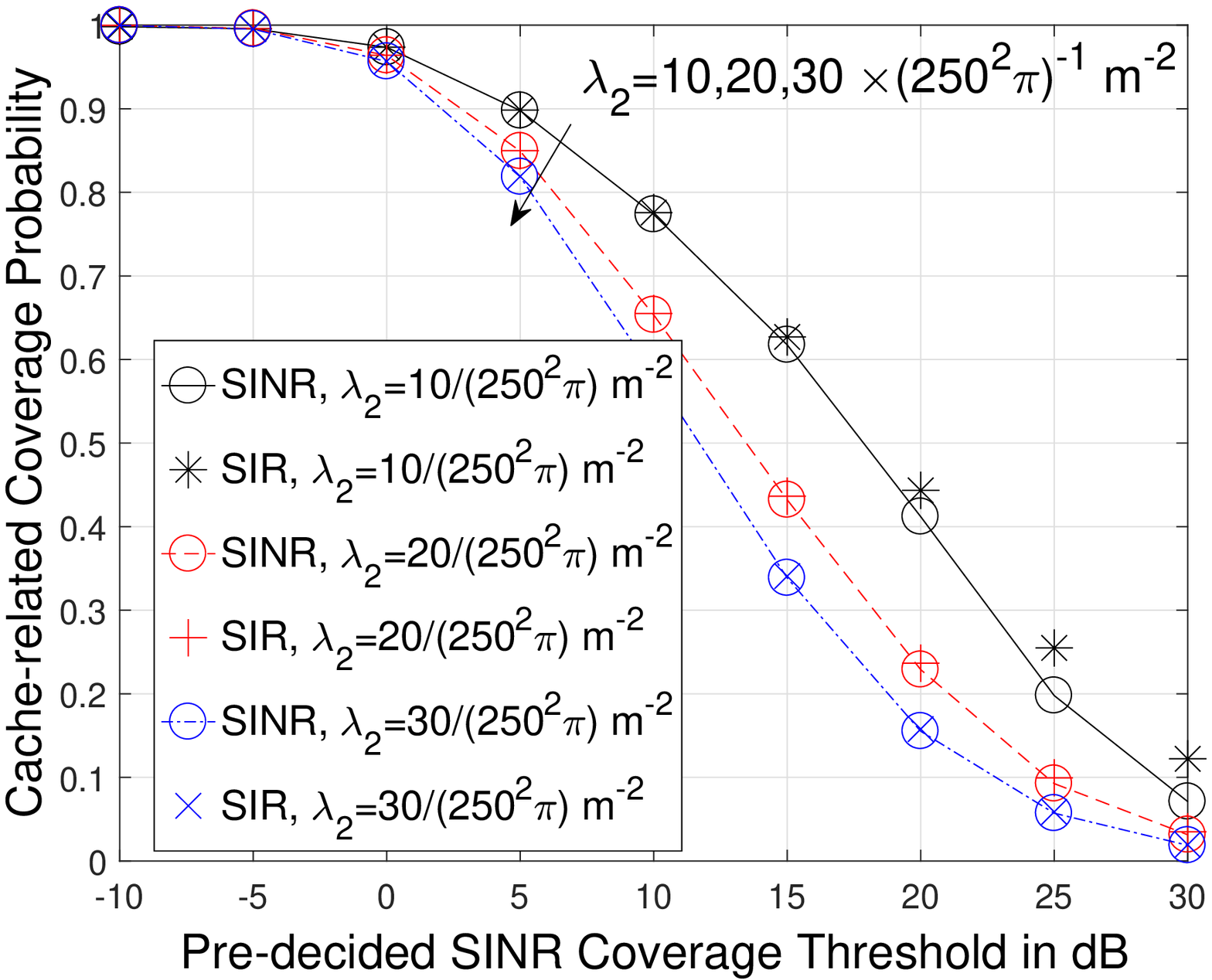}}
  \subfigure[Association probability with the second tier versus the density of the second tier $\lambda_2$, with $p_{i,f}=1$.]{\label{fig4} \includegraphics[width= 3.1in, height=2 in]{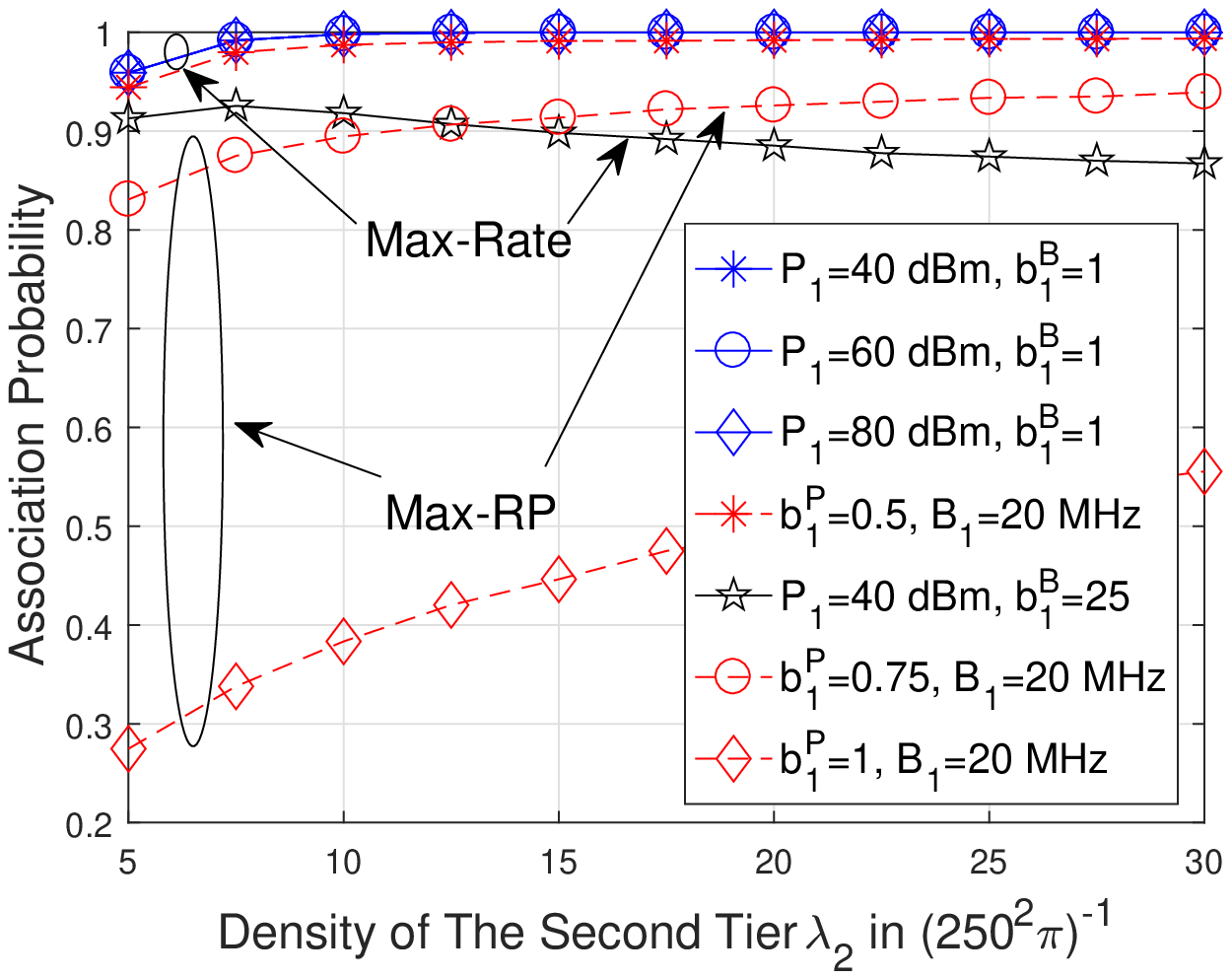}}
\caption{Interference-limited Property and The Impact of Antenna Scales.}
\end{figure*}

In this part, we corroborate the accuracy of our interference-limited assumption in dense mmWave networks. Fig.~\ref{fig3} shows the simulation results of SINR and SIR coverage probabilities with different $\lambda_2$. It can be seen from the figure that for $\lambda_2=20/(250^2\pi)$ m$^{-2}$ and $\lambda_2=30/(250^2\pi)$ m$^{-2}$, two ratios are almost overlapping and even at $\lambda_2=10/(250^2\pi)$ the difference between them is reasonably small. Accompanying with the condition $\lambda_1\ll \lambda_2$ assumed in Section~\ref{System Model}, we are able to confirm \textbf{Assumption~\ref{assumption1}} that the noise has a negligible impact on the coverage in the second tier due to high considered densities in our system. Therefore SINR can be approximated by SIR in the pico tier. As the first tier is also dominated by the interference rather than the noise as discussed in the previous part, we are capable of concluding that our cache-enabled hybrid HetNet is an interference-limited system.

\subsection{Impact of Transmit Power and Bias Factors}

The success probability is mainly decided by the association probability. Fig.~\ref{fig4} illustrates that the association probability under Max-Rate is independent of the transmit power of macro BSs as mentioned in \textbf{Remark~\ref{remark9}}. Regarding the bias factors, when the biased bandwidth $b_1^BB_1$ in Max-Rate changes from 500 MHz to 20 MHz, $(b_2^BB_2/b_1^BB_1)$ tends towards infinite and the association probability with the pico tier rises up to nearly one, which means under this condition, the typical UE always connects to the nearest pico BS as we discussed in \textbf{Remark~\ref{remark7}}. For Max-RP, the probability associating with the second tier has a negative correlation with $b_1^P$. Therefore, we are able to change the bias factors to balance the traffic load between two tiers as mentioned in \textbf{Remark~\ref{remark8}}.

\subsection{Impact of Antenna Scales and Cache Capacity}

\begin{figure*}[t!]
  \centering
   \subfigure[Success probability versus pre-decided rate threshold, with with $M_2=10$, $\lambda_2=20/(250^2\pi)$ m$^{-2}$, $C_{bh}=10^8$~bps, and $p_{i,f}=\textbf{U}(M_i-f)$.]{\label{fig5} \includegraphics[width= 3.1in, height=2  in]{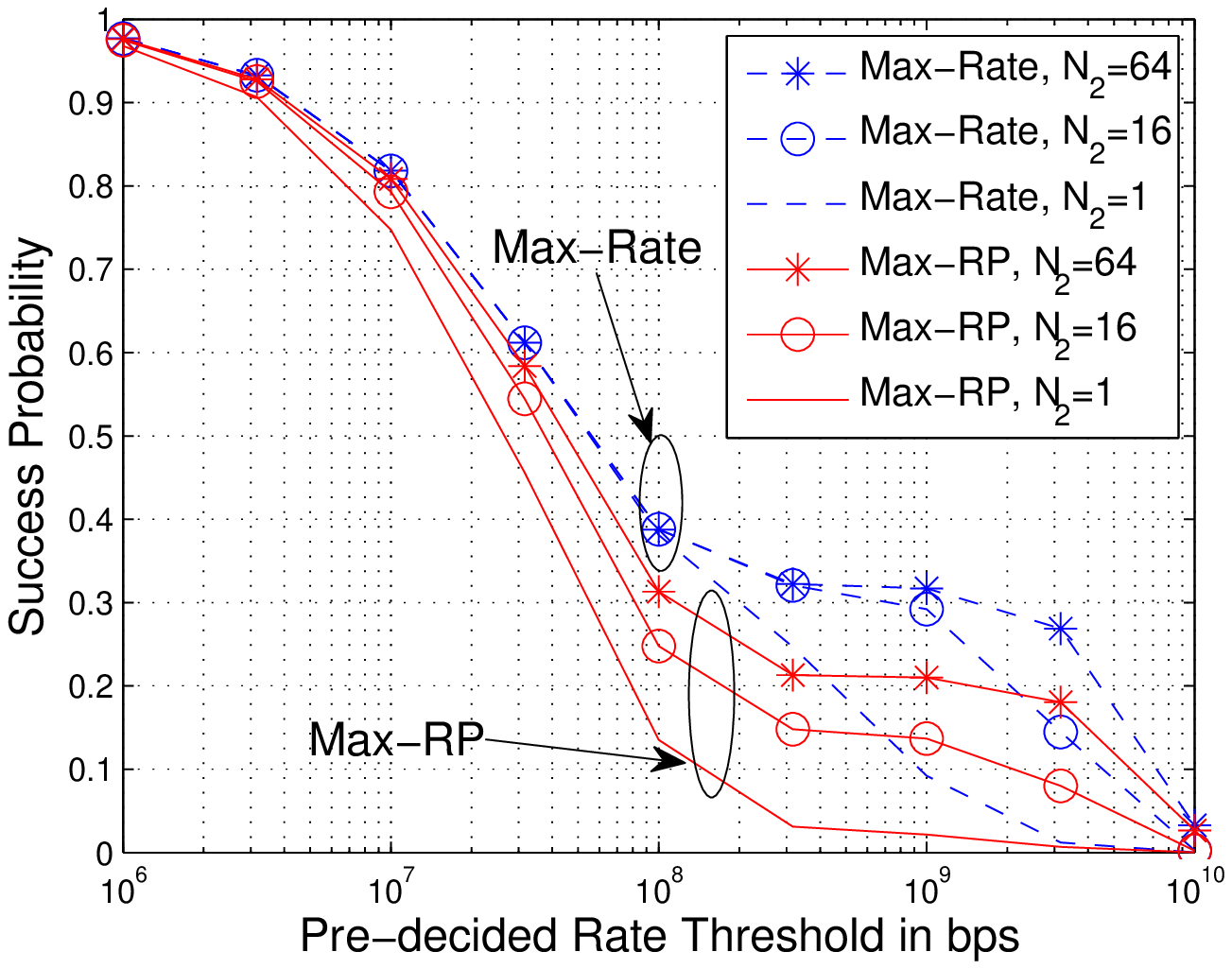}}
\subfigure[Success probability versus cache capacity $M_2$ in the second tier, with $R_{th}=10^8$ bps and $p_{i,f}=\textbf{U}(M_i-f)$.]{\label{fig6} \includegraphics[width= 3.1in, height=2  in]{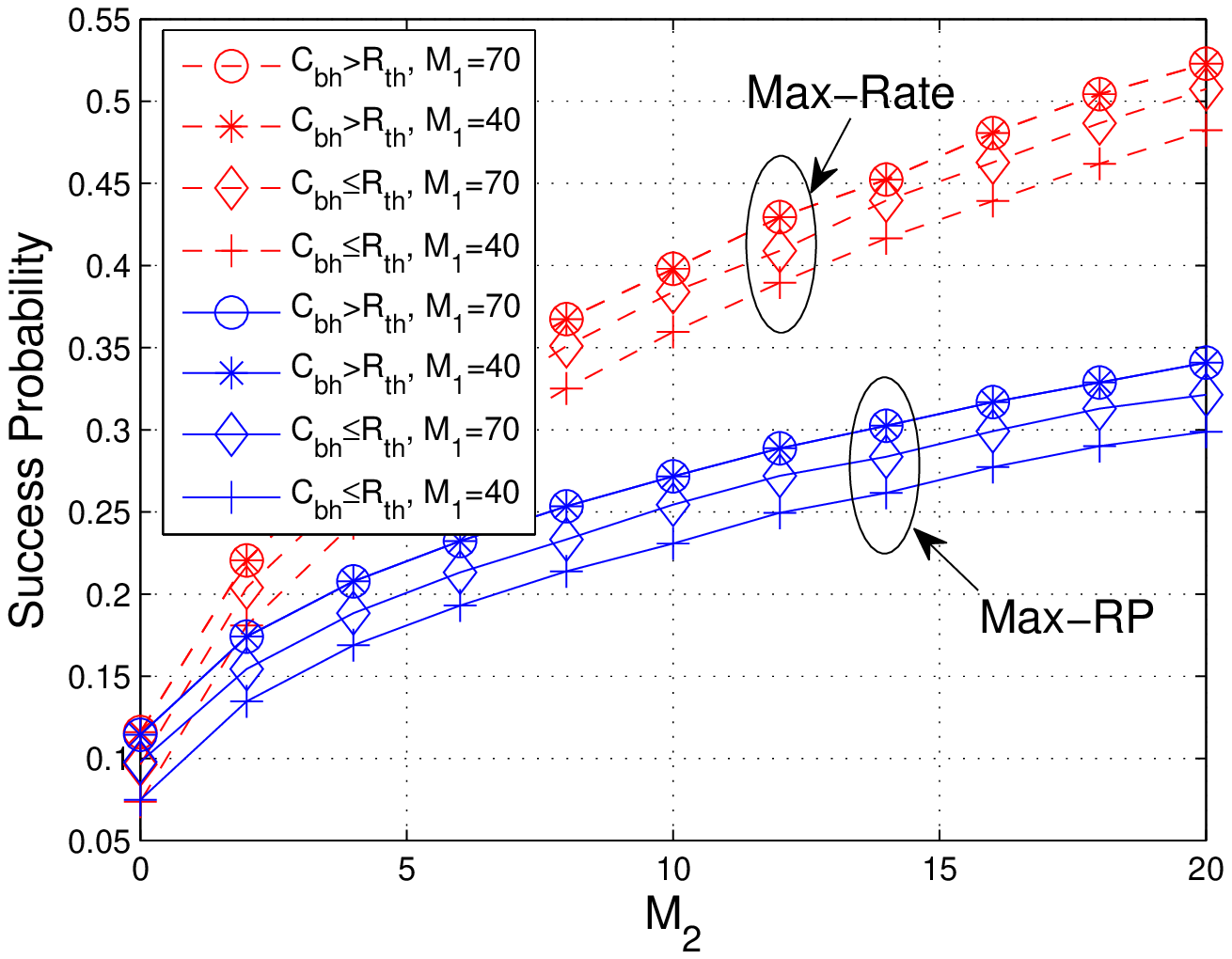}}
  \caption{Impact of Transmit Power and Bias Factors.}
\end{figure*}

As we adopt ULA with $N_2$ antenna elements at pico BSs, the increase of antenna scales $N_2$ enhances the received power and it also narrows the half-power beamwidth (HPBW). Additionally, the smaller HPBW contributes to the less interference in our system. Fig.~\ref{fig5} shows that the success probability arises with the increase of $N_2$, especially in the high pre-decided rate threshold $R_{th}$ region.

Cache capacity of BSs deployed in our networks is also an important parameter for analyzing the performance. The success probability is a monotonic increasing function with pico BSs' storage capacity $M_2$ as shown in Fig.~\ref{fig6}. Considering the cache capacity of macro BSs $M_1$, when $C_{bh}>R_{th}$, the success probability has no relationship with $M_1$ since the less-popular contents which are only contained in the server can be transmitted freely through backhauls. In this case, the proposed cache-enabled HetNet is same with the traditional one. On the other side, when $C_{bh}\leq R_{th}$, the server is blocked so that the success probability can be benefited by the large $M_1$, which represents that more multimedia files are stored at macro BSs.

\subsection{Performance of ASE and Various Carrier Frequencies}

\begin{figure*}[t!]
\centering
\subfigure[Area spectral efficiency versus pre-decided rate threshold, with $M_1=10$, $B_1=200$ MHz, and $p_{i,f}=\textbf{U}(M_i-f)$.]{\label{fig7} \includegraphics[width= 3.1in, height=2  in]{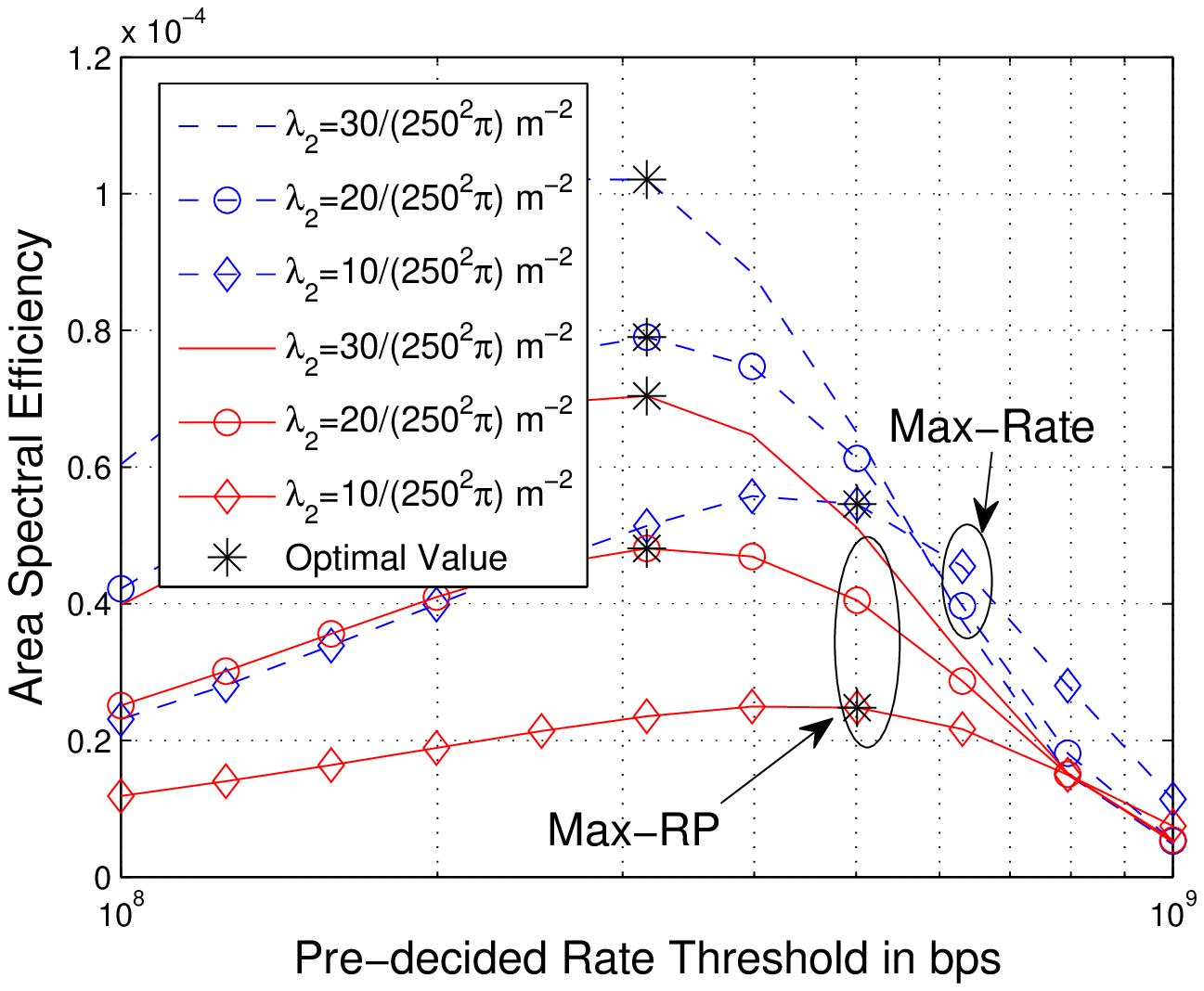}}
\subfigure[Success probability versus pre-decided rate threshold, with $M_2=20$ and $C_{bh}=10^8$ bps, and $p_{i,f}=\textbf{U}(M_i-f)$]{\label{fig8} \includegraphics[width= 3.1in, height=2  in]{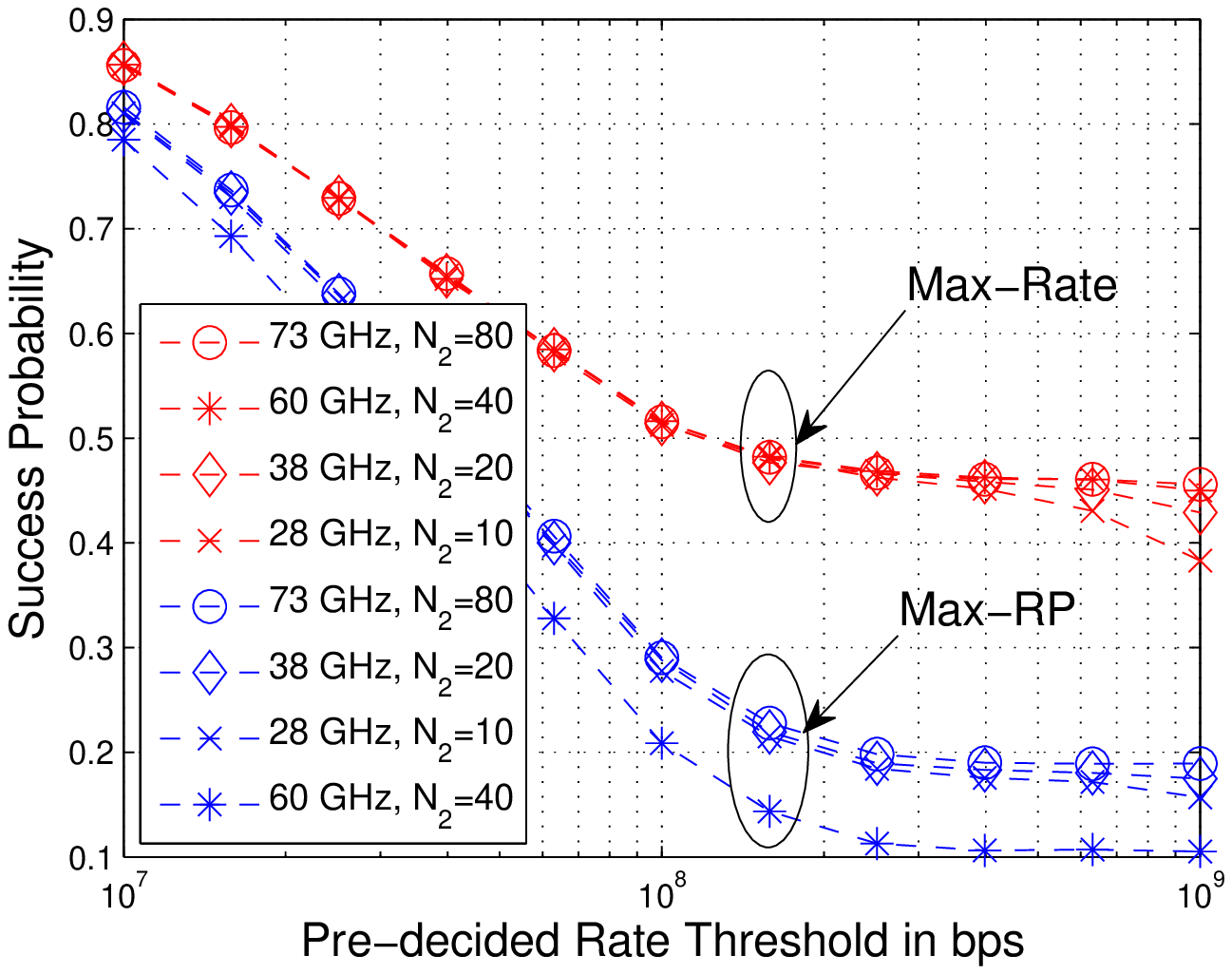}}
\caption{ASE and Various Carrier Frequencies.}
\end{figure*}
 We first present the performance of ASEs in Fig.~\ref{fig7}. It illustrates that Max-Rate scheme outperforms Max-RP scheme regarding the ASE. Moreover, the optimum pre-decided rate threshold $R_{th}$ for achieving the maximum ASE can be easily figured out from Fig.~\ref{fig7}, thereby corroborating \textbf{Remark~\ref{remark10}}. When the density of pico tier $\lambda_2$ increases from $10/(250^2\pi)$ m$^{-2}$ to $30/(250^2\pi)$ m$^{-2}$, the optimal value of $R_{th}$ decreases. In the real world, this optimum $R_{th}$ can be used to design a network with maximum ASE.
 \begin{table}[h]
\centering
\caption{Path Loss Exponent and Antenna Scales for The Second Tier}
\label{exponent}
\begin{tabular}{|c|c|c|c|c|}
\hline
   Carrier frequencies    & 28G    &38G    &60G     &73G\\ \hline
   Path loss exponent $\alpha_2$ for LOS links     & 2    &2   &2.25     &2  \\ \hline
   Number of antenna elements $N_2$  &10 &20 &40 &80 \\ \hline
\end{tabular}
\end{table}

 Based on the actual path loss exponents of LOS links~\cite{deng201528,rappaport201238} and estimated antenna scales~\cite{8016632} shown in Table~\ref{exponent}, the performance of four different carrier frequencies is illustrated in Fig.~\ref{fig8}. After comparing the carrier frequencies at 28 GHz, 38 GHz, 60 GHz, and 73 GHz, we conclude that 73 GHz is the best choice for both two user association strategies thanks to the largest antenna scales. Moreover, 28 GHz performs the worst among these four carrier frequencies in Max-Rate scheme due to the limited anti-interference ability, while 60 GHz causes the lowest success probability in Max-RP scheme because of the largest path loss exponent.

\subsection{Performance of Different Content Placement Policies}

\begin{figure}[ht]
  \centering
  \includegraphics[width= 3.1in, height=2 in]{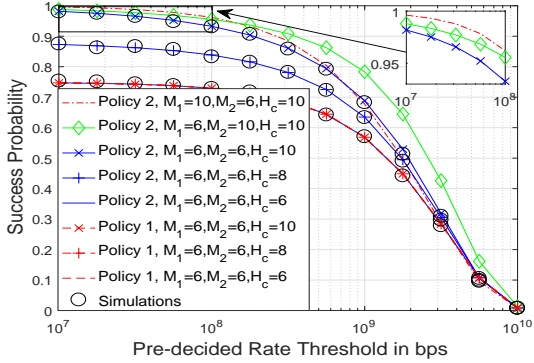}\\
  \caption{Success probability versus pre-decided rate threshold in Max-RP, with $B_1=500$ MHz, $C_{bh}=10^6$ bps, and $N_c=10$.}\label{comments}
\end{figure}
In the cache-enabled HetNet, the optimal content placement scheme is able to enhance the capacity of networks. Although the content distribution in the previous illustrations is a binary case, the proposed expressions in this work are suitable for other random distribution scenarios as shown in Fig.~\ref{comments}. We propose two content placement policies here for comparison: 1) policy 1: all BSs cache the most popular files, namely $p_{i,f}=\textbf{U}(M_i-f)$; and 2) policy 2: the files from the first to $H_c$ popular rank has the same probability of being cached at all BSs, namely $p_{i,f}=H_c/M_i$. For policy 1, the success probability is independent of $H_c$. However, the success probability in policy 2 has a positive correlation with $H_c$. Regarding the cache capacity, large $M_1$ for macro BSs slightly increases the success probability in low $R_{th}$ regions, while large $M_2$ for pico BSs significantly enhance the performance in high $R_{th}$ areas. The reason is that mmWave BSs is capable of providing faster data rate than sub-6 GHz BSs. When $M_1=M_2=H_c$, Policy 2 is same as policy 1, otherwise policy 2 outperforms policy~1.
\vspace{-0.3cm}
\section{Conclusion}

In this treatise, the performance of our cache-enabled hybrid HetNet has been analyzed in details. We have compared two different user association strategies with the aid of the stochastic geometry. More specifically, Max-Rate scheme outperforms Max-RP scheme regarding the success probability and ASE, but the difference between two association strategies can be eliminated by decreasing the transmit power of the macro tier. The proposed network, which performs better than the traditional HetNet, can be regarded as an interference-limited system due to the high density of mmWave tier and the nature of sub-6~GHz tier. We have analytically shown that the success probability of Max-Rate scheme is independent of $\lambda_1$, $N_1$, $C_i$ and $P_i$. Moreover, our system has a positive correlation with pico BSs' antenna scales and the cache capacity of both tiers. Additionally, there exists an optimum value of pre-decided rate threshold contributing to the maximum ASE. Lastly, for two user association strategies, 73 GHz is the best carrier frequency of mmWave tier. In addition to two considered content placement policies, our future work will focus on the optimization of the content placement schemes
\vspace{-0.3cm}
\numberwithin{equation}{section}
\section*{Appendix~A: Proof of Lemma~\ref{lemma1}} \label{appendixA}
\renewcommand{\theequation}{A.\arabic{equation}}
\setcounter{equation}{0}

The Laplace transform of interference in the second tier is given by (A.1) at the top of the next page.
\begin{figure*}[!t]
\normalsize
\begin{align}
\mathcal{L}_2( {s,\tau } ) =& \mathbb{E}\Big[ {\exp \big( { - ns\tau \sum\limits_{x \in {\Phi _2}\backslash x_0} {{G_2}( \omega  )|{h_2}{|^2}||x|{|^{ - {\alpha _2}}}} } \big)} \Big] \nonumber\\
\mathop  = \limits^{(a)} & {e^{ - 2\pi {p_{2,f}}{\lambda _2}{\mathbb{E}_{{G_2}}}\left[ {\int_r^{{R_L}} {(1 - {{(1 + \frac{{ns\tau {G_2}(\omega )}}{{N_2^p{v^{{\alpha _2}}}}})}^{ - N_2^p}})vdv} } \right] - 2\pi \left( {1 - {p_{2,f}}} \right){\lambda _2}{\mathbb{E}_{{G_2}}}\left[ {\int_0^{{R_L}} {(1 - {{(1 + \frac{{ns\tau {G_2}(\omega )}}{{N_2^p{v^{{\alpha _2}}}}})}^{ - N_2^p}})vdv} } \right]}} \nonumber\\
\mathop  = \limits^{(b)} &{e^{ - \frac{{\pi {\lambda _2}\lambda }}{d}\int_{ - \frac{d}{\lambda }}^{\frac{d}{\lambda }} {\left( {{p_{2,f}}\int_r^{{R_L}} {(1 - {{(1 + \frac{{ns\tau {G_2}(\omega )}}{{N_2^p{v^{{\alpha _2}}}}})}^{ - N_2^p}})vdv + \left( {1 - {p_{2,f}}} \right)\int_0^{{R_L}} {(1 - {{(1 + \frac{{ns\tau {G_2}(\omega )}}{{N_2^p{v^{{\alpha _2}}}}})}^{ - N_2^p}})vdv} } } \right)d\omega } }},
\end{align}
\hrulefill \vspace*{0pt}
\end{figure*}
For (A.1), (a) follows the Gamma random variable's moment generating function~\cite{6932503}; (b) is computing the expectation of the second tier antenna gain $G_2$.

When $\alpha_2>2$, (A.1) can be simplified into (A.2).
\begin{figure*}[!t]
\normalsize
\begin{align}
{\mathcal{L}_2}\left( {s,\tau } \right)\mathop  = \limits^{\left( c \right)} {e^{ - \pi {\lambda _2}(R_L^2 - {p_{2,f}}{r^2}) - \frac{{\pi {\lambda _2}\lambda }}{{2d}}\int_{ - \frac{d}{\lambda }}^{\frac{d}{\lambda }} {\left( {{p_{2,f}}S_2^0\left( {\frac{{X\left( \omega  \right)}}{{{r^{{\alpha _2}}}}}} \right){r^2} + \left( {1 - {p_{2,f}}} \right){\Delta _2}\left( {X\left( \omega  \right)} \right) - S_2^0\left( {\frac{{X\left( \omega  \right)}}{{R_L^{{\alpha _2}}}}} \right)R_L^2} \right)d\omega } }},
\end{align}
\hrulefill \vspace*{0pt}
\end{figure*}
For (A.2), $X\left( \omega  \right) = \frac{{s{G_2}\left( \omega  \right)\tau }}{{N_2^p}}$. (c) follows ${\Delta _i}(z) = \mathop {\lim }\limits_{r \to 0} {r^2}\left( {S_i^0(\frac{z}{{{r^{{\alpha _i}}}}}) - 1} \right)$ and Gauss hypergeometric function~\cite{liu2017caching}.

 When $\alpha_2=2$, (A.1) can be simplified into (A.3).
 \begin{figure*}[!t]
\normalsize
\begin{align}
 {\mathcal{L}_2}\left( {s,\tau } \right)\mathop  = \limits^{\left( d \right)}{e^{ - \pi {\lambda _2}(R_L^2 - {p_{2,f}}{r^2}) - \frac{{\pi {\lambda _2}\lambda }}{{2d}}\int_{ - \frac{d}{\lambda }}^{\frac{d}{\lambda }} {\left( {X\left( \omega  \right)\left( {{F_y}\left( {\frac{{X\left( \omega  \right)}}{{R_L^2}}} \right) - {p_{2,f}}{F_y}\left( {\frac{{X\left( \omega  \right)}}{{{r^2}}}} \right)} \right)} \right)d\omega } }}.
\end{align}
\hrulefill \vspace*{0pt}
\end{figure*}
For (A.3), (d) follows (2.117-1), (2.117-3) and (2.118-1) in~\cite{jeffrey2007table}. With the aid of  Gauss-Chebyshev Quadrature, we obtain \textbf{Lemma~\ref{lemma1}}. The proof is finished.
\vspace{-0.3cm}
\section*{Appendix~B: Proof of Corollary~\ref{corollary2}} \label{appendixB}
\renewcommand{\theequation}{B.\arabic{equation}}
\setcounter{equation}{0}

Under the \textbf{Assumption~\ref{assumption1}}, we ignore the noise effect. The PDF of the coverage probability in the second tier is shown as follows
\begin{align}
{p_{{\Upsilon _{2,f}}}}( \tau  )&= - \frac{{d}}{{d\tau }}\mathbb{P}[ {{\Upsilon _{2,f}} > \tau } ] \nonumber \\
&\mathop\approx \limits^e \sum\limits_{n = 1}^{N_2^p} {{{( - 1)}^{n + 1}}{N_2^p \choose n}\int_0^{{R_L}} {{{ f}_D}(r,\tau )} } dr,
\end{align}
where $\mathbb{P}[ {{\Upsilon _{2,f}} > \tau } ]={{ P}_{{\Upsilon _{2,f}}}}( \tau  )$ in \textbf{Corollary~\ref{corollary1}} and ${{ f}_D}( {r,\tau } )$ is the derivative of $-{ F_D}( r,\tau )$. (e) follows the fact that variable $\tau$ is only contained in ${ F_D}( r,\tau )$. Then ${{ f}_D}( {r,\tau } )$ is given by
\begin{align}
{{ f}_D}( {r,\tau } ) =-{P_{2,f}}( r ) \frac{{d}}{{d\tau }}\mathcal{L}_2\big( {\frac{{{n\eta _L}{r^{{\alpha _2}}}}}{{{G_0}}}},\tau \big).
\end{align}

With the aid of \textbf{Lemma~\ref{lemma2}}, we obtain
\begin{align}
{{ f}_D}(r,\tau ) \mathop  =  {{ F}_D}( {r,\tau } )\frac{{\pi {\lambda _2}\lambda }}{{d}}\int_{ - \frac{d}{\lambda }}^{\frac{d}{\lambda }}{w_f}(\omega ,r,\tau )d\omega.
\end{align}

By substituting (B.3) into (B.1) and then applying Gauss-Chebyshev Quadrature, we obtain \textbf{Corollary~\ref{corollary2}}. The proof is finished.
\vspace{-0.3cm}
\section*{Appendix~C: Proof of Lemma~\ref{lemma3}} \label{appendixC}
\renewcommand{\theequation}{C.\arabic{equation}}
\setcounter{equation}{0}

The derivative of the coverage probability under Max-Rate scheme can be calculated with the coverage probabilities of two tiers discussed in Section~\ref{Coverage_Probability}. We first figure out the probability of the first tier coverage based on Max-Rate strategy as follows
\begin{align}
&F_{1,f }^R( \tau  ) \nonumber \\
=& \mathbb{P}\big[ {{\Upsilon_{1,f}} > \tau |b_1^B{B_1}{{\log }_2}( {1 + {\Upsilon_{1,f}}} ) > b_2^B{B_2}{{\log }_2}( {1 + {\Upsilon_{2,f}}} )} \big]\nonumber\\
=& \int_\tau ^\infty  {{p_{{\Upsilon _{1,f}}}}( {{\Upsilon_{1,f}}} )} \mathbb{P}\left[ {{\Upsilon_{2,f}} < {{( {1 + {\Upsilon_{1,f}}} )}^{\frac{{b_1^B{B_1}}}{{b_2^B{B_2}}}}} - 1} \right]d{\Upsilon_{1,f}}\nonumber\\
=& \int_\tau ^\infty  {{p_{{\Upsilon _{1,f}}}}( {{\Upsilon_{1,f}}} )} ( {1 - {P_{{\Upsilon _{2,f}}}}( {{{( {1 + {\Upsilon_{1,f}}} )}^{\frac{{b_1^B{B_1}}}{{b_2^B{B_2}}}}} - 1} )} )d{\Upsilon_{1,f}}.
\end{align}

Then, the PDF of such coverage probability $f_{1,f}^R( \tau  )$ is given by
\begin{align}
f_{1,f}^R( \tau  )=& - \frac{d}{{d\tau }} { F_{1,f}^R( \tau  )} \nonumber\\
 =&-\frac{d}{{d\tau }}\int_\tau ^\infty  {{p_{{\Upsilon _{1,f}}}}( {{\Upsilon_{1,f}}} )}\nonumber \\
 &\times( {1 - {P_{{\Upsilon _{2,f}}}}( {{{( {1 + {\Upsilon_{1,f}}} )}^{\frac{{b_1^B{B_1}}}{{b_2^B{B_2}}}}} - 1} )} )d{\Upsilon_{1,f}}\nonumber\\
=&{p_{{\Upsilon _{1,f}}}}( {{\tau }} )( {1 - {P_{{\Upsilon _{2,f}}}}( {{{( {1 + {\tau }} )}^{\frac{{b_1^B{B_1}}}{{b_2^B{B_2}}}}} - 1} )} ).
\end{align}

Using the same method, the PDF for the probability of the second tier coverage under Max-Rate scheme is shown as below
\begin{align}
f_{2,f}^R( \tau  ) =  {p_{{\Upsilon _{2,f}}}}( {{\tau }} )( {1 - {P_{{\Upsilon _{1,f}}}}( {{{( {1 + {\tau }} )}^{\frac{{b_2^B{B_2}}}{{b_1^B{B_1}}}}} - 1} )} ).
\end{align}
Combing (C.2) and (C.3), we obtain \textbf{Lemma~\ref{lemma3}}. The proof is finished.
\vspace{-0.5cm}
\bibliographystyle{IEEEtran}
\bibliography{mybib}
\end{document}